\documentclass[acmsmall,screen]{acmart}

\usepackage{multirow, listings}
\usepackage{hyperref}
\usepackage{pifont}
\usepackage{colortbl} % Required for \rowcolor
\usepackage{xcolor} % Required for defining custom colors
\usepackage{subcaption}
\usepackage{amsmath,amsfonts} % math
\usepackage{pgfplots}
\usepackage{graphicx}
\usepackage{tcolorbox}
\usepackage{soul}
\usepackage{wrapfig}
\usepackage{xspace}
\usepackage{ulem}
\usepackage[T1]{fontenc}
\usepackage{array}
\usepackage{adjustbox}
\usepackage{multirow}
\usepackage{makecell}
\usepackage[inline]{enumitem} % RQ enumeration
\usepackage{tikz} % For drawing shapes
\usepackage{calc} % For inline images
\usepackage{mwe} % for placeholder image

\usepackage{algpseudocode}  % backwards-compatible algorithmic environment
\usepackage{algorithm}  % floating environment for algorithms

\usepackage{tcolorbox} % used for evaluation result box
\usepackage{tabularx}  % used for benchmark programs table
\usepackage{booktabs}  % used for evaluation tables
\usepackage{multirow}  % used for eval table
\usepackage{url}  % used for links
\usepackage{ulem} %
\usepackage{tabularx}

\setstcolor{red}

\algrenewcommand\algorithmicrequire{\textbf{Input:}}
\algrenewcommand\algorithmicensure{\textbf{Output:}}

\newcommand{\PreserveBackslash}[1]{\let\temp=\\#1\let\\=\temp}
\newcolumntype{C}[1]{>{\PreserveBackslash\centering}p{#1}}
\newcolumntype{R}[1]{>{\PreserveBackslash\raggedleft}p{#1}}
\newcolumntype{L}[1]{>{\PreserveBackslash\raggedright}p{#1}}

\newcommand*{\algrule}[1][\algorithmicindent]{\hspace*{.5em}\vrule\vrule
width 0pt height \baselineskip depth .25\baselineskip\hspace*{\dimexpr#1-.5em}}
\makeatletter
\newcount\ALG@printindent@tempcnta
\def\ALG@printindent{%
    \ifnum \theALG@nested>0% is there anything to print
    \ifx\ALG@text\ALG@x@notext% is this an end group without any text?
    \else
    \unskip
    \ALG@printindent@tempcnta=1
    \loop
    \algrule[\csname ALG@ind@\the\ALG@printindent@tempcnta\endcsname]%
    \advance \ALG@printindent@tempcnta 1
    \ifnum \ALG@printindent@tempcnta<\numexpr\theALG@nested+1\relax% can't do <=, so add one to RHS and use < instead
    \repeat
    \fi
    \fi
}%
\usepackage{etoolbox}
\patchcmd{\ALG@doentity}{\noindent\hskip\ALG@tlm}{\ALG@printindent}{}{\errmessage{failed to patch}}
\makeatother
\algtext*{EndWhile}% Remove "end while" text
\algtext*{EndFor}% Remove "end for" text
\algtext*{EndIf}% Remove "end if" text
\algblockdefx[ForB]{ForB}{EndForB}[0]{\textbf{for} }{$)$}
\algcblockdefx[DoB]{ForB}{DoB}{EndDoB}{\textbf{do}}{\textbf{end for}}
\algtext*{EndDoB}% Remove "end for" text
\newcommand{\header}[1]{\cellcolor{black}\textcolor{white}{#1}}
\newcommand*{\circled}[2][]{\tikz[baseline=(C.base)]{
    \node[inner sep=0pt] (C) {\vphantom{1g}#2};
    \node[draw, fill=gray!25, circle, inner sep=1pt, yshift=1pt] 
        at (C.center) {\vphantom{1g}};
    \node[inner sep=0pt] (D) {\vphantom{1g}#2};
}}

\newcommand{\tool}{\textsc{TargetFuzz}}

\definecolor{forestgreen}{HTML}{228B22}
\definecolor{orangep}{rgb}{0.71, 0.43, 0.89}
\definecolor{orp}{rgb}{1, 0.5, 0.2}
\definecolor{dkgreen}{rgb}{0,0.6,0}
\definecolor{gray}{rgb}{0.3,0.3,0.3}
\definecolor{mauve}{rgb}{0.58,0,0.82}
\definecolor{PastelBlue}{HTML}{BBCCEE}
\definecolor{PastelGreen}{HTML}{CCDDAA}
\definecolor{PastelOrange}{HTML}{EEEEBB}
\definecolor{PastelRed}{HTML}{FFCCCC}
\definecolor{PastelPurple}{HTML}{CCEEFF}
\definecolor{PastelCyan}{HTML}{CCEEFF}
\definecolor{GreyHL}{rgb}{0.9,0.9,0.9}

\definecolor{colorold}{HTML}{BD613C}
\definecolor{colornew}{HTML}{6D8C00}%92a65f 
\definecolor{coloremph}{HTML}{dcde9f}

\definecolor{NodeBlue}{HTML}{66B2FF}
\definecolor{NodeGreen}{HTML}{66FF66}
\definecolor{NodeRed}{HTML}{FF6666}
\definecolor{NodeYellow}{HTML}{FFFF66}
\definecolor{hg}{HTML}{CCDDAA} % highlight green
\definecolor{hb}{HTML}{CCEEFF} % highlight blue
\definecolor{hr}{HTML}{FFCCCC} % highlight red
\definecolor{hy}{HTML}{FFF2CC} % highlight yellow
\definecolor{hp}{HTML}{E1D5E7} % highlight purple

\lstdefinestyle{CStyle}{
    language=C,
    basicstyle=\ttfamily\scriptsize,  % Changed to \scriptsize
    keywordstyle=\color{blue},
    commentstyle=\color{green!50!black}\ttfamily,
    stringstyle=\color{red},
    numbers=left,
    numberstyle=\tiny\color{gray},
    stepnumber=1,
    numbersep=5pt,
    showstringspaces=false,
    tabsize=2,
    breaklines=true,
    breakatwhitespace=false,
    frame=single,
    escapeinside={||}
}

\lstdefinestyle{PseudoStyle}{
    language=,
    basicstyle=\ttfamily\scriptsize,  % Changed to \scriptsize
    commentstyle=\color{green!50!black}\ttfamily,
    morekeywords={if, then, else, while, for, return},
    keywordstyle=\color{blue},
    numbers=left,
    numberstyle=\tiny\color{gray},
    stepnumber=1,
    numbersep=5pt,
    showstringspaces=false,
    tabsize=2,
    breaklines=true,
    breakatwhitespace=false,
    frame=single,
    escapeinside={||}
}

\lstdefinestyle{ANTLRStyleNew}{
    frame=tb,
    rulecolor=\color{black},
    basicstyle=\color{blue}\scriptsize\ttfamily, % Use small monospace font
    commentstyle=\color{green!50!black}\ttfamily, % Comments in green italics
    breaklines=true, % Allow line breaks
    tabsize=2, % Set tab width to 2 spaces
    moredelim=[s][\color{green!50!black}\ttfamily]{'}{'}, % Single quotes in green
    moredelim=*[s][\color{black}\ttfamily]{options}{\}}, % 'options' in black until trailing '}'
    morecomment=[l]{//}, % Single-line comments
    morecomment=[s]{/*}{*/}, % Multi-line comments
    emph={%
        STRING%                                            literal strings listed here
        },emphstyle={\color{blue}\ttfamily},%              and formatted in blue
    alsoletter={:,|,;,+,?,*,\(,\)}, % Treat ':', '|', ';', '+', '?', '*', '(', ')' as keywords
    morekeywords={:,|,;,+,?,*,\(,\)}, % Define these special characters
    morekeywords={grammar, fragment, returns}, % ANTLR-specific keywords
    keywordstyle={\color{purple}}, % Format these characters in black
}

\lstdefinestyle{MLIRStyle}{
    language=C,
    basicstyle=\ttfamily\scriptsize,  % Changed to \scriptsize
    keywordstyle=\color{blue},
    commentstyle=\color{green!50!black}\ttfamily,
    stringstyle=\color{red},
    numbers=left,
    numberstyle=\tiny\color{gray},
    stepnumber=1,
    numbersep=5pt,
    showstringspaces=false,
    tabsize=2,
    breaklines=true,
    breakatwhitespace=false,
    frame=single,
    escapeinside={||},
    morekeywords={func, arith, vector, constant, extract, return}, % Add MLIR-specific keywords
}

\lstdefinestyle{mypythonstyle}{
    frame=tb,
    language=Python,
    basicstyle=\color{black!}\scriptsize\ttfamily, % Use small monospace font
    commentstyle=\color{green!50!black}\ttfamily, % Comments in green italics
    breaklines=true, % Allow line breaks
    tabsize=2, % Set tab width to 2 spaces
    morecomment=[l]{//}, % Single-line comments
    morecomment=[s]{/*}{*/}, % Multi-line comments
    stringstyle=\color{red!80!black}, 
    emph={%
        STRING%                                            literal strings listed here
        },emphstyle={\color{blue}\ttfamily},%              and formatted in blue
    alsoletter={:,|,;,+,?,*,\(,\)}, % Treat ':', '|', ';', '+', '?', '*', '(', ')' as keywords
    keywordstyle={\color{blue}}, % Format these characters in black
    keywords=[2]{balanced,FOR\_STMT\_,WHILE_STMT_,ARITH_EXPR_}, % Group 2: variables
    keywordstyle=[2]\color{brown}, 
    keywords=[3]{Construct},  % Group 3: control flow keywords
    keywordstyle=[3]\color{cyan},% Style for Group 3 keywords
    keywords=[4]{\[,\]},  % Group 3: control flow keywords
    keywordstyle=[4]\color{cyan},% Style for Group 3 keywords
}

\newlength\myheight
\newlength\mydepth
\settototalheight\myheight{Xygp}
\newcommand*\inlinegraphics[1]{%
  \settototalheight\myheight{Xygp}%
  \settodepth\mydepth{Xygp}%
  \raisebox{-\mydepth}{\includegraphics[height=\myheight]{#1}}%
}

\newtcolorbox{resultbox}{
    colback=black!5!white,
    colframe=white!50!black, 
    boxsep=0mm,
}

\usetikzlibrary{fit,calc}

\newcommand{\llvmtriggerimprv}{2.8$\times$}
\newcommand{\mlirtriggerimprv}{2.6$\times$}

\newcommand{\llvmbrcovimprv}{8\%}
\newcommand{\mlirbrcovimprv}{11\%}

\newcommand{\llvmnumoptim}{37}
\newcommand{\llvmbettergm}{26 / \llvmnumoptim}
\newcommand{\llvmavgtriggerimprvgm}{1.74$\times$}

\newcommand{\llvmbettergrayc}{23 / \llvmnumoptim}
\newcommand{\llvmavgtriggerimprvgrayc}{5.14$\times$}

\newcommand{\llvmworsesynth}{22 / \llvmnumoptim}
\newcommand{\llvmavgtriggerimprvsynth}{1.64$\times$}

\newcommand{\llvmtfvalidity}{41.14\%}
\newcommand{\llvmsynthvalidity}{19.18\%}

\newcommand{\mlirnumoptim}{26}
\newcommand{\mlirnumdialects}{7}
\newcommand{\mliravgtriggerimprvgm}{2.71$\times$}
\newcommand{\mlirbettergm} {22 / \mlirnumoptim}

\newcommand{\mlirbettersynth} {18 / \mlirnumoptim}
\newcommand{\mliravgtriggerimprvsynth}{2.45$\times$}

\newcommand{\loc}{5360}
\newcommand{\locCAnnot}{323}
\newcommand{\locMlirAnnot}{314}

\newcommand{\graycOptimModuleLineCoverage}{12.12\%}
\newcommand{\graycOptimPassLineCoverage}{8.66\%}
\newcommand{\graycOptimPassUncoveredRatio}{47.5\%}

\newcommand{\numbugs}{18}

\newcommand{\nummiddlebugs}{10}
\newcommand{\numbackendbugs}{2}

\newcommand{\numcrash}{14}
\newcommand{\numwrongcode}{4}

\begin{document}

\title{Targeted Testing of Compiler Optimizations via Grammar-Level Composition Styles}

% ===== Authors =====
\author{Zitong Zhou}
\affiliation{%
  \institution{University of California, Los Angeles (UCLA)}
  \city{Los Angeles}
  \state{CA}
  \country{USA}
}
\email{zitongzhou@cs.ucla.edu}

\author{Ben Limpanukorn}
\affiliation{%
  \institution{University of California, Los Angeles (UCLA)}
  \city{Los Angeles}
  \state{CA}
  \country{USA}
}
\email{blimpan@cs.ucla.edu}

\author{Hong Jin Kang}
\affiliation{%
  \institution{The University of Sydney}
  \city{Sydney}
  \state{NSW}
  \country{Australia}
}
\email{hongjin.kang@sydney.edu.au}

\author{Jiyuan Wang}
\affiliation{%
  \institution{Tulane University}
  \city{New Orleans}
  \state{LA}
  \country{USA}
}
\email{wjiyuan@tulane.edu}

\author{Yaoxuan Wu}
\affiliation{%
  \institution{University of California, Los Angeles (UCLA)}
  \city{Los Angeles}
  \state{CA}
  \country{USA}
}
\email{thaddywu@cs.ucla.edu}

\author{Akos Kiss}
\affiliation{%
  \institution{University of Szeged}
  \city{Szeged}
  \country{Hungary}
}
\email{akiss@inf.u-szeged.hu}

\author{Renata Hodovan}
\affiliation{%
  \institution{University of Szeged}
  \city{Szeged}
  \country{Hungary}
}
\email{hodovan@inf.u-szeged.hu}

\author{Miryung Kim}
\affiliation{%
  \institution{University of California, Los Angeles (UCLA)}
  \city{Los Angeles}
  \state{CA}
  \country{USA}
}
\email{miryung@cs.ucla.edu}
% ===================


\begin{abstract}

Ensuring the correctness of optimizing compilers is critical yet challenging. Fuzzing offers a promising path, but existing fuzzers struggle to test optimization logic effectively. In our preliminary study, the state-of-the-art LLVM fuzzer GrayC covered only 12\% of LLVM’s optimization module—and left 47\% of optimization passes entirely untested—after four hours of coverage-guided fuzzing. We identify two key challenges. First, most fuzzers use optimization pipelines (heuristics-based, fixed sequences of passes) as their harness. Because of the phase-ordering problem, pass ordering can enable or preempt transformations, so pipelines inevitably miss many optimization interactions; moreover, many optimizations are not scheduled, even at aggressive levels. Second, optimizations typically fire only when inputs satisfy specific structural relationships, which existing generators and mutations struggle to produce.

We propose targeted fuzzing of individual optimizations to complement pipeline-based testing. Our key idea is to exploit composition styles—structural relations over program constructs (e.g., adjacency, nesting, repetition, ordering)—that optimizations look for. We build a general-purpose, grammar-based mutational fuzzer, \tool{}, that (i) mines candidate composition styles from an optimization-relevant corpus, then (ii) rebuilds those styles inside different contexts offered by a larger, generic corpus via lightweight, synthesized mutations to exercise variations of optimization logic. \tool{} can be easily adapted to a new programming language by providing lightweight, grammar-based, construct annotations—and it automatically synthesizes mutators and crossovers to rebuild composition styles. There is no need for hand-coded generators or language-specific mutators, which is particularly useful for modular frameworks such as MLIR, whose dialect-based, rapidly evolving ecosystem makes optimizations difficult to fuzz. Our evaluation on LLVM and MLIR shows that \tool{} improves coverage by \llvmbrcovimprv{} and \mlirbrcovimprv{} and triggers optimizations \llvmtriggerimprv{} and \mlirtriggerimprv{}, compared to baseline fuzzers under the targeted fuzzing mode. We show that targeted fuzzing is complementary: it effectively tests all 37 sampled LLVM optimizations, while pipeline-fuzzing missed 12.

\end{abstract}


\setcopyright{none} % to remove the copyright notice
\settopmatter{printacmref=false} % to remove the ACM Reference Format
\renewcommand\footnotetextcopyrightpermission[1]{}

\maketitle

\section{Introduction}
Modern optimizing compilers are monolithic software, composed of layers that transform a high-level program into machine instructions. As optimizing compilers evolve, their growing complexity introduces new challenges for ensuring correctness \cite{llvmPasses,mlirPasses}. Optimizing compilers can crash, hang, or miscompile, where bugs silently make their way into production software. In an empirical study of over 10K confirmed optimization bugs in GCC and LLVM, Zhou et al. reported that the optimizer is one of the most buggy components in the  compilers \cite{zhou2021optimbugs}.
Sun et al.~ conducted a systematic study of compiler bugs in GCC and LLVM and found that high priority tends to be assigned to optimizer bugs, most notably 30\% of the bugs in GCC’s inter-procedural analysis component are labeled to be the highest priority (P1)~\cite{sun2016compilerbugs}. 
Given the importance of optimizing compilers in the software supply chain, ensuring their correctness is essential to the reliability of virtually all software. 

Fuzzing or automated random testing offers a practical way to expose these hidden flaws in compilers. 
However, existing fuzzers are limited in their ability to thoroughly test compilers' optimization components. We performed artifact evaluation of the state-of-the-art C-language fuzzer GrayC~\cite{mendoza2023grayc}. Using its default configuration of feeding input programs to test LLVM through \texttt{clang -O3}, after 4 hours of fuzzing, \graycOptimPassUncoveredRatio\ of the optimization passes were entirely untested. 
It only achieves total \graycOptimModuleLineCoverage\ line coverage in the transformation module of LLVM. On source files of optimization passes (excluding utilities), it has only \graycOptimPassLineCoverage\, line coverage on average.

Most fuzzers (like GrayC) drive the compiler through aggressive, canonical optimization pipelines (e.g., \texttt{clang -O3}) in hopes of maximizing optimization behaviors. There are 2 key challenges to this approach. First, the classic \emph{phase-ordering problem}~\cite{Vegdahl1982phasecoupling,Benitez1988phaseOrdering,Whitfield1997gospel}: modern optimizers apply optimization passes by heuristics, so the order of passes can enable or preempt one another. An example in LLVM is that "(fully) unrolling loops early removes opportunities for the loop vectorizer"~\cite{Meijer2021dontUnrollBeforeVectorisation}.
In practice, this means that in fuzzing, many optimizations are overshadowed by others and subsequently under-tested, as the GrayC experiment demonstrates.
Second, even aggressive pipelines are not \textit{exhaustive} as they need to balance compile time and stability, so not every optimization is enabled by default~\cite{Meijer2021loopPassesDisabled}. In addition, the growing popularity of modular and extensible compiler frameworks exacerbates the problem. For example, MLIR~\cite{mlir-lang-ref} allows developers to build custom IR, passes and pipelines, making it even more difficult to exercise many optimizations in a single canonical pipeline. 
To address these two challenges, we propose to \textit{complement whole-pipeline fuzzing (good for end-to-end interactions) with \textbf{targeted pass fuzzing} that exercises corner-case optimization interactions which the pipelines rarely reach.} We expect targeted fuzzing to isolate and test individual optimization passes or small groups of passes, by feeding inputs directly to them, therefore addressing these challenges.
We detail the limitations of whole-pipeline fuzzing in Section~\ref{sec:background} and targeted pass fuzzing in Section~\ref{sec:approach}.

Another fundamental challenge to fuzzing compiler optimizations is that fuzzer-generated inputs cannot easily satisfy the triggering conditions expected by optimizations. This is because triggering optimizations requires further structural constraints on program inputs. For instance, optimization \textbf{Loop Fusion} (i.e., combining loop bodies) only triggers when two immediately adjacent loops share the same trip count \cite{IBM2019loopfusionslides}. 
\textit{We observe that compiler optimizations target many structural relations (such as adjacency, repetition, and nesting) on grammar-based structures.} For example, LLVM's \textbf{Loop Flatten}~\cite{Meijer2018loopflatten} flattens nested loops; and \textbf{Loop Fusion} fuses adjacent loops. Our intuition traces back to the theoretical foundations of compiler optimizing transformations \cite{allen1971catalogue, Aho1970formal2optims}. That is, conceptually, an optimizing transformation consists of a pattern-matcher and a rewrite rule \cite{mlirPatternRewriter}. The pattern-matcher finds structural relations expressed on intermediate representations or abstract syntax trees. We propose a taxonomy of such structural relations defined at the level of grammar-based constructs, coupled with  mutator synthesis to reconstruct these optimization-triggering relations on a large corpus. In this paper, we refer to these structural relations as {\em composition styles}.

For example, we define the \textsf{Balanced} composition that can match arithmetic expressions located in multiple branches of an \texttt{if-then-else} construct. \textsf{Balanced} maps to a set of composition-specific mutations, including {Replicate}, which, given a new program, tries to locate a similar \texttt{if-then-else} construct containing an arithmetic expression, then replicates that expression in other branches, thus reconstructing the \textsf{Balanced} style in a new context. The idea is that by reconstructing these optimization-triggering relations in a variety of code contexts, we can exercise diverse variations of optimization logic.

We embody this idea of composition-based mutational fuzzing in a tool named \tool{}. 
The core functionality of \tool{} is that it defines a set of composition styles such as \textsf{Balanced}. For each composition style, a compiler developer can \textit{register} it to match relevant \textit{program constructs} that compiler optimizations operate on, such as if-else blocks, loops, logical and arithmetic expressions, function calls, arrays, memory references. Program constructs serve as a lightweight representation that \tool{} understands and operates on, unifying different programming languages. The only cost to target a programming language with \tool{} is to define program constructs with respect to the grammar of the language, which we estimate to take a few hours—considerably cheaper than the effort of building custom test generators.

Our contributions are as follows:
\begin{enumerate}
    \item We highlight the limitations of fuzzing optimizing compilers using canonical pipelines, and propose targeted pass fuzzing to complement it. Our evaluation shows that targeted fuzzing can effectively reach optimizations that are otherwise under-tested.
    \item We are the first to 1) formalize a taxonomy of structural relations (composition styles) that compiler optimizations target, and to 2) propose automatically-synthesized mutations and crossovers that rebuild these relations, for the purpose of mutational fuzzing.
    \item We implement \tool{} that embodies this idea of composition-based mutational fuzzing. \tool{} by design supports both targeted fuzzing and pipeline fuzzing. We show in evaluation that \tool{} improves coverage by \llvmbrcovimprv,~\mlirbrcovimprv~and triggers optimizations \llvmtriggerimprv,~\mlirtriggerimprv, compared to baseline fuzzers under targeted fuzzing mode. \tool{} achieves higher coverage than all grammar-based fuzzers, and comparable coverage to the language-specific custom fuzzer GrayC, in whole-pipeline fuzzing mode. In addition, we show that targeted fuzzing is complementary to pipeline fuzzing: it effectively tests all 37 sampled LLVM optimizations, while pipeline-fuzzing missed 12. 
\end{enumerate}

\section{Background} \label{sec:background}

Compiler fuzzers typically use optimization pipelines (e.g., \texttt{-O3}) as the test harness, feeding test programs directly into the pipeline and observing for crashes and miscompilations. 
This is convenient to configure, but it systematically leaves significant portions of the optimization space untested for two reasons. This motivates the need for \emph{targeted pass fuzzing}.
\vspace{-0.2em}
\paragraph{(Challenge 1A) Phase ordering problem.}
Optimizers typically implement a large number of optimization passes (>200 for GCC, LLVM, MLIR~\cite{gccPasses,llvmPasses,mlirPasses}). Contrary to superoptimizers~\cite{massalin1987superoptimizer,Schkufza2013stochsuperoptimizer,Bansal2006AutoPeepholeSuperopt} that aim to enumerate a large space of optimization sequences and select the optimal sequence, mainstream, production-quality optimizers build pipelines that are fixed, empirically-proven sequences of passes, in order to speed up compilation. Critically, within these heuristics-based pipelines, the order of passes can easily enable or preempt opportunities for one another, known as optimization interactions/coupling. For example, loop unrolling may remove opportunities for subsequent loop optimizations~\cite{Meijer2021dontUnrollBeforeVectorisation}. Decades of work show that a fixed, optimal order for all programs does not exist, and searching for a good sequence is a difficult problem~\cite{Vegdahl1982phasecoupling,Benitez1988phaseOrdering,Whitfield1997gospel,Kulkarni2003findingoptsequence,Kulkarni2004fastsearchPhaseSeq,Triantafyllis2003compilerOptSpaceExploration,Purini2013goodOptSequence,Kulkarni2012mitigatephaseOrdering}. Consequently, no pipeline can exercise every transformation path: some opportunities are created only after specific prior passes; others are destroyed by them. In fuzzing, this manifests as certain \emph{cold} optimizations that are rarely tested under the pipeline.
\vspace{-0.2em}
\paragraph{(Challenge 1B) Optimization pipelines do not include all passes.}
Modern compilers design their default pipelines to balance compile time and stability rather than ``run every pass.'' 

Many transformations are \emph{off by default} due to profitability or maturity concerns. For example, several loop transforms have historically been disabled by default, to avoid regressions until cost models and implementations mature~\cite{Meijer2021loopPassesDisabled}. This challenge is exacerbated as the compiler ecosystem is shifting from monolithic optimizers toward \emph{modular, extensible} infrastructures like MLIR. It organizes IR into \emph{dialects} that each captures a different abstraction and employs different optimizations~\cite{mlir-lang-ref}. It is widely adopted in machine-learning and domain-specific compilers such as CIRCT~\cite{circt}, Triton~\cite{triton}, Mojo~\cite{mojo}, and OpenXLA~\cite{openxla}. By design there is no single canonical sequence that covers all dialects and optimizations. Compiler fuzzers of tomorrow must adapt to this modularity.

\vspace{-0.2em}
\paragraph{(Challenge 2) Optimizations expect structural relations.} Besides the 2 challenges brought by whole-pipeline fuzzing, we identified another fundamental, orthogonal challenge: optimizations are guarded by a series of strict structural and semantic conditions, making them hard to test through fuzzing. In modern compilers, conceptually, each optimization pass consists of (1) an analysis phase that checks whether structural and semantic conditions are satisfied, and (2) a transformation phase that optimizes the program.
We use \textbf{Loop Fusion} in LLVM as an example. Loop Fusion is an optimization pass that merges multiple loops into one, to reduce overhead in loop control structures and achieve better instruction-level parallelism \cite{IBM2019loopfusionslides}. It is important in data-intensive applications and machine learning compilers. In Figure~\ref{fig:loopfusion-example-listing}, the optimizer attempts to apply Loop Fusion on the input program on the left. Loop Fusion iteratively checks that each condition is satisfied. \textbf{Adjacency} requires that the two loops have no instructions in between; \textbf{Conformance} requires that the two loops have the same trip counts. There are also other semantic conditions that must be satisfied such as profitability. 
\textit{If any condition is unsatisfied, the pass stops early, and the critical transformation logic is not executed.} In Figure~\ref{fig:loopfusion-example-listing}, the input program satisfies all conditions, producing the optimized program with a fused loop.
In this paper, we refer to such structural relationships between program constructs such as loops as \emph{composition styles}. We present the formal definitions in Section~\ref{sec:approach}. While useful composition styles exist in fuzzing corpora, to the best of our knowledge, no existing fuzzers explicitly leverage them to test optimizations.

\begin{figure}[tb]
    \centering
    \begin{minipage}[t]{0.45\textwidth}  % Adjusted width
        \centering
\begin{lstlisting}[style=CStyle]
int sum1=0, sum2=0;
for(int i=0;i<10;i++){
	sum1+=i;
} // adjacent loops
for(int j=0;j<10;j++){
	sum2+=j;
}
\end{lstlisting}
        % \subcaption{A test program that triggers Loop Fusion.}
        \label{fig:loopfusion-example-before}
    \end{minipage}
    \hfill
    \begin{minipage}[t]{0.45\textwidth}  % Adjusted width
        \centering
\begin{lstlisting}[style=CStyle]
int sum1=0,sum2=0;
for(int i=0;i<10;i++){
	sum1+=i;
	sum2+=i; // body of loop 2 moved here
}
\end{lstlisting}
        % \subcaption{The optimized program after Loop Fusion.}
        \label{fig:loopfusion-example-after}
    \end{minipage}
    \caption{A test program that triggers Loop Fusion and its optimized output.}
    \label{fig:loopfusion-example-listing}
    \vspace{-1em}
\end{figure}

\vspace{-0.2em}
\paragraph{Goal.}
We set out to design a general-purpose fuzzer that \emph{exploits optimization-specific composition styles} to effectively test optimizing compilers. The fuzzer must be \emph{lightweight}: the user only needs to supply a small corpus from where the composition styles are extracted. It must also be \emph{easily adaptable} across compilers and languages (e.g., MLIR compilers), so that porting requires only supplying a grammar plus concise annotations---no need to hand-code custom generators or mutators. It must support \emph{both} targeted pass fuzzing and whole-pipeline fuzzing.

\section{Approach}
\label{sec:approach}
\begin{figure*}[t]
    \centering
    \includegraphics[width=0.95\linewidth]{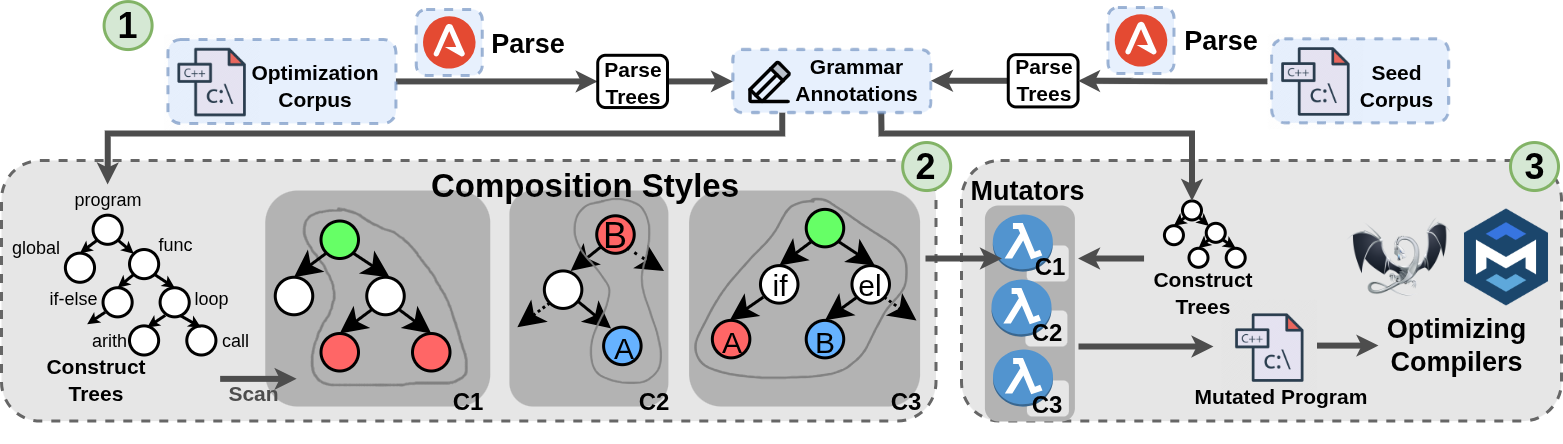}
    % \caption{Phase 1: \tool{} extracts composition styles, e.g., loops are adjacent, from the optimization corpus - programs that contain 'breadcrumbs' of what's necessary to trigger optimizations. Phase 2: each composition style synthesizes mutators, e.g. replicating a loop. Mutators are applied on the seed corpus to reconstruct compositions and test optimizations }
    \caption{Phase 1: \tool{} uses the supplied grammar to parse the optimization corpus and seed corpus into grammar parse-trees, then it uses supplied grammar annotations to translate them respectively into construct trees. Phase 2: \tool{} extracts composition styles, e.g., loops are adjacent, from the optimization construct trees - programs that contain 'breadcrumbs' of what's necessary to trigger optimizations. Phase 3: each composition style synthesizes mutators, e.g. replicating a loop. Mutators are applied on the seed construct trees to reconstruct compositions and test optimizations. Blue boxes are user inputs.}
    \label{fig:flowchart}
    \vspace{-0.75em}
\end{figure*}

% We propose \tool{}, a grammar-based, mutational fuzzing framework that synthesizes mutators to target optimizations by recognizing and amplifying code composition styles.
Figure~\ref{fig:flowchart} gives an overview of \tool{}'s fuzzing pipeline. In each iteration, from a target-specific corpus program, \tool{} extracts a composition style that may trigger an optimization. 
Then, it synthesizes {\em targeted mutators} and applies them to a recipient program to rebuild the composition style.
% \hj{In the previous version, we had the "overview" diagram and I thought that was pretty helpful to explain the overall pipeline? }
% ~\jw{100\% agree. A workflow figure that includes all four inputs can be super helpful to explain how \tool{} works.}

\tool{} takes \textit{an optimization corpus, a seed corpus, a grammar, and grammar annotation as inputs.} The \textit{optimization corpus} is \emph{where the composition styles are extracted}. It consists of programs that are relevant to target optimizations. The assumption is that this corpus already has the breadcrumbs for composition styles necessary to trigger optimizations. It can be sourced from the compiler's unit tests, generated by Large Language Models (LLMs) or manually written by compiler developers. Note that \tool{} does not require the optimization corpus to be perfect, i.e., a program does not necessarily need to satisfy all conditions and trigger the optimization; it just needs to have the \emph{breadcrumbs}. 
\textit{The optimization corpus enables \tool{} to support two fuzzing modes in a single tool:} (1) targeted fuzzing mode for testing individual optimizations (e.g., \texttt{-loop-fusion}), where the optimization corpus only needs to contain a handful of programs that are relevant to the targeted optimization. This is particularly valuable for MLIR, e.g., to fuzz optimizations of the \texttt{async} dialect (abstractions of asynchronous execution), the optimization corpus can be \texttt{async}'s unit tests;
and (2) whole pipeline fuzzing to trigger all kinds of optimizations through an aggressive pipeline (e.g., \texttt{clang -O3}), when we pool optimization corpora. In evaluation, we show the two fuzzing modes are complementary to each other in terms of code coverage.

The \textit{seed corpus} serves a different purpose. While the optimization corpus provides hints for exercising optimizations, the seed corpus provides the overall diversity of inputs. It is a generic corpus of programs that can be sourced from compiler benchmarks like SPEC CPU 2017~\cite{speccpu207} or produced by existing program generators like CSmith~\cite{yang2011csmith}, GrayC~\cite{mendoza2023grayc}, YarpGen~\cite{Livinskii2020yarpgen}.

The \textit{grammar} is a context-free grammar that defines the syntax of the programming language and is used to parse programs from both corpora. The \textit{grammar annotation} (detailed in Section~\ref{sec:constructs}) extends the grammar by marking specific production rules that correspond to \textit{program constructs} relevant to compiler optimizations (e.g., \textsc{Vector}, \textsc{FuncCall}), which may not be directly identifiable from the raw grammar rules alone. Program constructs also serve as a representation that unifies the syntax of different programming languages, enabling \tool{} to be language-agnostic.

\subsection{Program Constructs} \label{sec:constructs}
% \ez{briefly go over ANTLR grammar and how annotation for matching constructs work, add 2 code listings}

\begin{figure}[tb]
    \centering
    \begin{minipage}[t]{0.38\textwidth}  % Adjusted width
        \centering
\begin{lstlisting}[style=ANTLRStyleNew]
/** C11 Spec */
grammar C;

addExpr
	: mulExpr (('+' | '-') mulExpr)*;
statement
	: labeledStatement
	| iterationStatement
	| ...;
iterationStatement
	: forStatement
	| whileStatement
	| doWhileStatement;
\end{lstlisting}
        \subcaption{A snippet of the ANTLR C grammar with three production rules. $'|'$ separates alternatives of each rule.}
        \label{lst:antlr_c}
    \end{minipage}
    \hfill
    \begin{minipage}[t]{0.58\textwidth}  % Adjusted width
        \centering
\begin{lstlisting}[style=mypythonstyle]
FOR_STMT_ = Construct(["forStatement"])
LOOPS_ = CmpdConstruct([FOR_STMT_, WHILE_STMT_, ...])
def isarith(node):
	return node.name == "addExpr" and \
		(any([n.src in ["+", "-", "*", ...] for n in lexers(node)]) \
		or is_numerical(node))
ARITH_EXPR_ = CustomConstruct(isarith)

def is_literal_array(node):
  ...
LTR_ARR_ = CustomConstruct(is_literal_array)
VECTOR_EXPR_ = CmpdConstruct([LTR_ARR_, ARR_ACES_])
\end{lstlisting}
        \subcaption{\tool{}'s \texttt{Construct} API enables defining custom program constructs that are relevant to compiler optimizations, through \textit{grammar annotation}.}
        \label{lst:gram_annot}
    \end{minipage}
    % \begin{minipage}[t]{0.50\textwidth}  % Adjusted width
    %     \centering
    %     \lstinputlisting[style=mypythonstyle]{Program/balanced.py}
    %     \subcaption{\tool{} registers \texttt{balanced} composition on annotated constructs.}
    %     \label{lst:register_balanced}
    % \end{minipage}
    \begin{minipage}[t]{0.99\textwidth}  % Adjusted width
        \centering
        \includegraphics[trim={1.2cm 0 10cm 0},clip,width=0.99\linewidth]{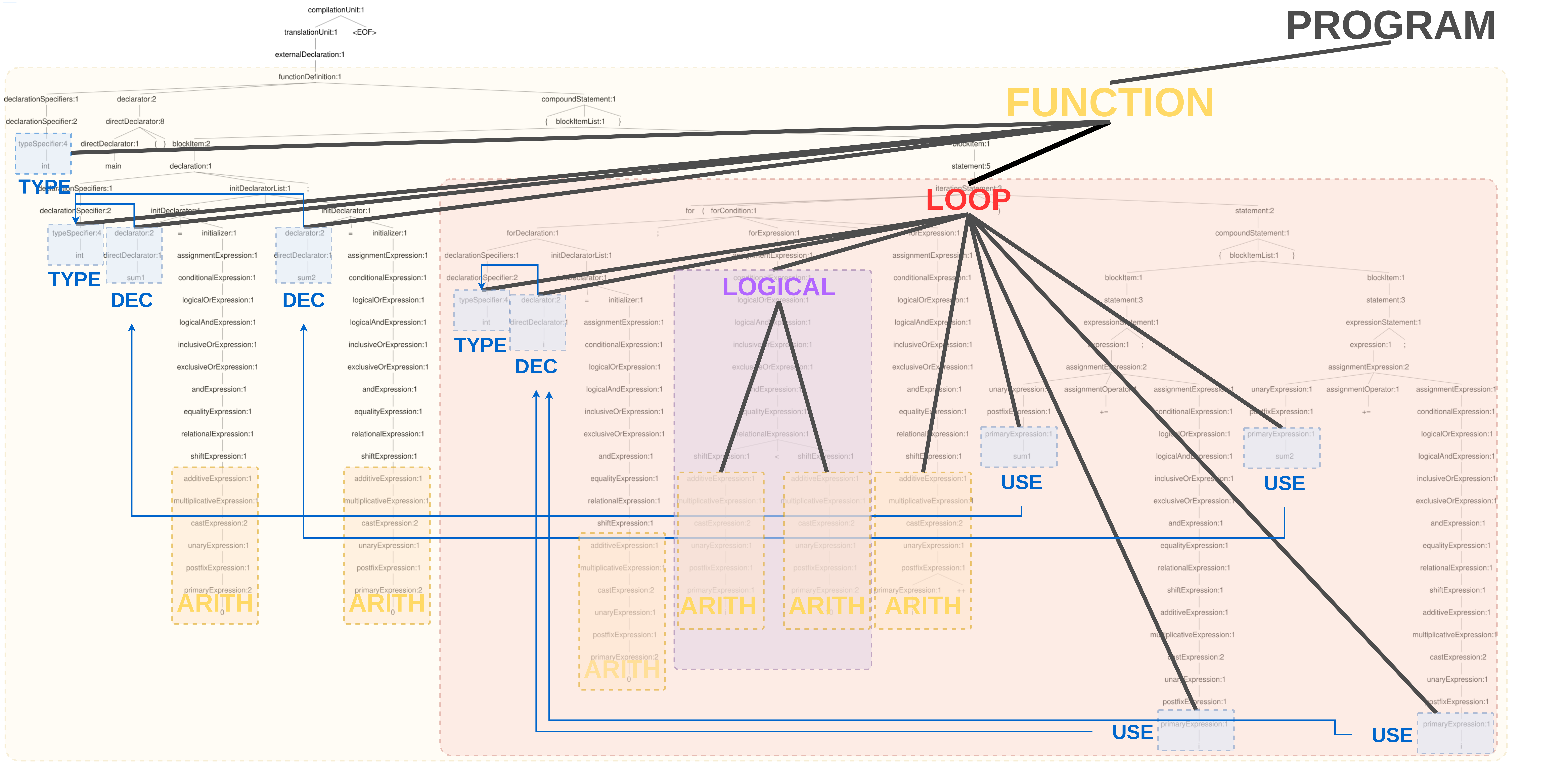}
        \subcaption{Construct Tree of the optimized C program in Figure~\ref{fig:loopfusion-example-listing}. Grammar parse-tree shown in the background. Colored boxes are nodes (constructs), bold lines are edges. Blue arrows are type-declaration-use chains.}
        \label{fig:construct_tree}
    \end{minipage}
    \vspace{-.5em}
    \caption{Listing~\ref{lst:antlr_c} is C grammar. Listing~\ref{lst:gram_annot} defines its program constructs. Listing~\ref{fig:construct_tree} shows construct tree.}

    % Listing~\ref{lst:register_balanced} registers the \textsf{Balanced} composition.
    \label{fig:program_constructs}
    \vspace{-1em}
\end{figure}

In Phase 1 of Figure~\ref{fig:flowchart}, \tool{} uses an \inlinegraphics{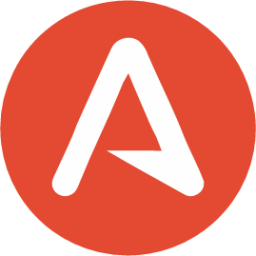} ANTLR context-free grammar to parse both the optimization and seed corpus \cite{antlrwebsite}. 
% The core of an ANTLR grammar is parser production rules. 
% For example, in Listing~\ref{lst:antlr_c}, the C grammar snippet contains three production rules. Rule \texttt{addExpr} defines that additive expressions consist multiplicative expressions  \texttt{mulExpr} connected by additive operators (\texttt{+} or \texttt{-}).
% The rules \texttt{statement} and \texttt{iterationStatement} can represent multiple subrules.
Since ANTLR grammars are designed for generating parsers rather than pattern-matching, 
they do not always contain the production rules for semantic program constructs that compiler optimizations look for. 
For example, while the C grammar includes rules such as \texttt{forStatement} \& \texttt{ifStatement}, that match for-loop and if-block, respectively, it lacks the rules for fine-grained constructs such as vectors, function calls, and memory references. 
The postfix expression rule \texttt{postfixExpr} can represent all three aforementioned constructs, but it does not distinguish them. This is problematic because compiler optimizations often operate on these fine-grained constructs.
% \ben{I think saying it "lacks the rules" for these constructs implies that the grammar cannot parse these constructs. Since we're going with the arith expression example in the next paragraph, I think it would be better to discuss it here too. Something like: While certain constructs like if statements and for statements exactly map to specific production rules in the C grammar (i.e. ifStatement and forStatement), other constructs like arithmetic expressions only map to a production rule if it satisfies certain conditions.}
% \hj{would be helpful if we also explained at a high-level what this custom construct is used for.}

\tool{} allows practitioners to \textit{define custom program constructs by annotating any programming language's grammar} using its \texttt{Construct} API.
For example, Listing~\ref{lst:gram_annot} illustrates four program constructs: \texttt{FOR\_STMT\_} (\textsc{ForLoop}), \texttt{LOOPS\_} (\textsc{Loop}), \texttt{ARITH\_EXPR\_} (\textsc{Arithmetic}), and \texttt{VECTOR\_EXPR\_} (\textsc{Vector}); the last two are not represented by rules in the grammar and need to be custom defined. \texttt{FOR\_STMT\_} is defined as any node in the grammar parse-tree of the rule \texttt{forStatement}. \texttt{LOOPS\_} is a compound construct that is either a \texttt{FOR\_STMT\_}, \texttt{WHILE\_STMT\_}, or \texttt{DO\_WHILE\_STMT\_}.
\texttt{ARITH\_EXPR\_} is a custom construct defined by the predicate function \texttt{isarith}. It is a \texttt{addExpr} grammar node (additive expressions) that is either numerical or contains arithmetic operators. Similarly, \texttt{VECTOR\_EXPR\_} is a compound construct of \texttt{LTR\_ARR} (literal array) and \texttt{ARR\_ACES\_} (array access), both custom constructs defined using a predicate function.

% \kim{this paragraph seems in conflict with the previous paragraph, where we say the C grammar includes rules such as forStatement and ifStatement. Do you think it's better to change the in-text example to discuss MemRef  or Vector instead?} \ez{I added \texttt{Vector}. Even if the grammar includes production rules like \texttt{forStatement} that map directly to constructs, we still need to explicitly define them \texttt{FOR\_STMT\_ = Construct("FOR\_STMT\_", ["forStatement"])}}~\jw{Agree with Miryung. If we still need to define a for statement, then the sentence in the previous paragraph, "while the C grammar includes rules such as forStatement
% \& ifStatement, ...it lacks the rules for fine-grained constructs such as vectors..", should add "besides, even if they have forStatement in the grammar, they are not applicable because xxx", and justify why define them \texttt{FOR\_STMT\_ = Construct("FOR\_STMT\_", ["forStatement"]} can solve xxx}

Program constructs also serve as a lightweight, unified representation that enables \tool{} to be programming language-agnostic, e.g. construct \texttt{ARITH\_EXPR\_} represents arithmetic expressions universally. 
Conceptually, as shown in Figure~\ref{fig:flowchart}, \tool{} \emph{translates} grammar parse-tree into a \emph{construct tree}, where each node is a program construct. Figure~\ref{fig:construct_tree} visualizes this step, the optimized C program in Figure~\ref{fig:loopfusion-example-listing} is parsed into a grammar parse-tree that is shown in the background, overlayed by the construct tree. Each colored box is a construct node, and bold lines are edges. Blue arrows are declaration-use chains that \tool{} tracks for scope well-formedness.

When adapting \tool{} to fuzz a new programming language $\mathcal{L}$, developers need to annotate $\mathcal{L}$'s grammar as in Figure~\ref{lst:gram_annot} to map specific grammar parse-tree nodes to the corresponding internal constructs like \texttt{ARITH\_EXPR\_}.
\tool{} defined common program constructs for C and MLIR, such as \textsc{Loop}, \textsc{If-Else}, \textsc{FuncCall}, \textsc{MemRef}, \textsc{Vector}, \textsc{Arithmetic}, and \textsc{Logical}, \textsc{Jump}, and many more, using 323 and 314 lines of grammar annotations, respectively. To employ \tool{} on a new language, practitioners need to annotate its grammar with relevant constructs, which typically means few hours of effort.

\subsection{Composition Styles} \label{sec:pattern}
% Define a command for drawing a circle with a number inside

\newcommand{\cellcontent}[1]{\raisebox{0pt}[8pt][8pt]{#1}}
            
\begin{table*}[tb]
    \scriptsize
    \centering
    % \scalebox{0.99}{
    % \definecolor{headercolor}{RGB}{150,50,50} % Defines the color used for the header
    \begin{tabular}{|>{\sffamily}l|>{\sffamily}p{6.5cm}|>{\sffamily}p{2.2cm}|>{\sffamily}p{.7cm}|p{1.7cm}|
    }
        \header{Styles} & \header{Program Constructs Types $\langle \mathcal{T}_1,\dots,\mathcal{T}_k\rangle$} & \header{Context Types $\mathcal{T}_{\text{ctx}}$ } & \header{Mutations} & \header{Predicates} \\ \toprule
        
        \textbf{Cousins}  & \textbf{$T_1 = T_2 \in \mathcal{T}$}: $\big\{$\textsc{Loop}, \textsc{FuncCall}, \textsc{Arithmetic}, \textsc{Logical} $\big\}$ & $\big\{$\textsc{If-Else}, \textsc{Loop}, \textsc{Func}$\big\}$ & ALL & $k$:generation distance, $d$:token distance\\ \hline
        
        \textbf{Nesting} & \textbf{$T_1 = T_2 \in \mathcal{T}$}: $\big\{$\textsc{Loop}, \textsc{If-Else} $\big\}$ & $\big\{$\textsc{Loop}, \textsc{Func}$\big\}$ & ALL & $d$:nesting depth\\ \hline
        
        \textbf{Precedes}   & \textbf{$T_{l}, T_{r}\in\mathcal{T}$}: $\big\{$\textsc{FuncCall}, \textsc{MemRef}, \textsc{Arithmetic}$\big\}$ & $\big\{$\textsc{Loop}, \textsc{Func}$\big\}$ & \circled{M} \circled{I} \circled{P} & \\ \hline
        
        \textbf{Balanced}  & \textbf{$T_1=$} \textsc{If-Else}; \textbf{$T_2=T_3 \in$} $\big\{$\textsc{Loop}, \textsc{FuncCall}, \textsc{Arithmetic}, \textsc{MemRef} $\big\}$ & $\big\{$\textsc{If-Else}, \textsc{Loop}, \textsc{Func}$\big\}$ & ALL & $d$:branch depth\\ \hline
        
        \textbf{Sequence} & \textbf{$T_1 = \dots = T_n \in \mathcal{T}$}: $\big\{$\textsc{Loop}, \textsc{FuncCall}, \textsc{Vector}, \textsc{Arithmetic} $\big\}$ & $\big\{$\textsc{Loop}, \textsc{Func}$\big\}$ & \circled{R}\circled{I}\circled{P} &$l$:seq. length\\ \hline
        
        \textbf{Exists} & \textbf{$T \in \mathcal{T}$}: $\big\{$\textsc{Func}, \textsc{Loop}, \textsc{If-Else}, \textsc{FuncCall}, \textsc{\textsc{MemRef}}, \textsc{Vector}, \textsc{Arithmetic}, ...$\big\}$ & $\big\{$\textsc{Vector}, \textsc{Loop}, \textsc{If-Else}, \textsc{Func}, \textsc{Program}$\big\}$ & \circled{I}\circled{P} &$l$:min tokens\\ \hline
   % \bottomrule
    \end{tabular}
    % }
    \vspace{1ex}
    \caption{\tool{}'s instantiated composition styles. 
    For example, \textbf{Precedes} matches 2 constructs $T_l$ and $T_r$ of type \textsc{FuncCall}, \textsc{MemRef}, or \textsc{Arithmetic}, and a common ancestor construct (context) $T_\text{ctx}$ of type \textsc{Loop} or \textsc{Func}. It is equipped with three mutators: \circled{M}ove, \circled{I}nsert, and \circled{R}eplace (See Section~\ref{sec:mutator}).}
    \label{tab:patterns}
    \vspace{-2.5em}
\end{table*}

\newcommand{\imagewidth}{0.45\linewidth}

\begin{figure*}[tb]
    \centering
    \begin{minipage}{\linewidth}
        \centering
        \begin{subfigure}{\imagewidth}
            \centering
            \includegraphics[width=\linewidth]{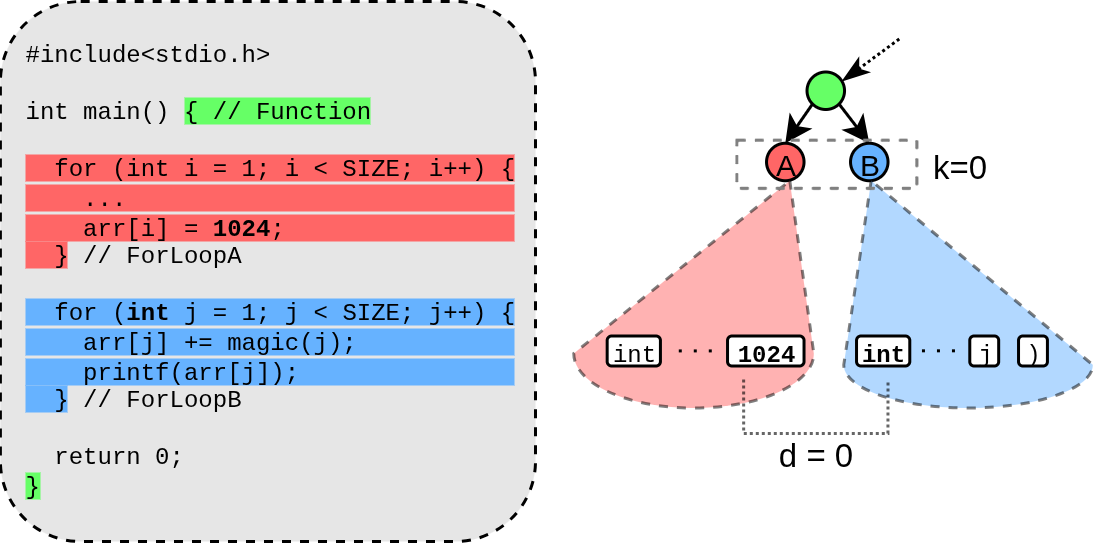}
            \caption{
            {\sffamily\textbf{Cousins}} matches \texttt{ForLoopA}, \texttt{ForLoopB} in context \texttt{Function}. %Parameters: deg. of separation $k = 2$, distance $d = 0$.
            }
            \label{fig:cousins-example}
        \end{subfigure}
        \hspace{2em} % Adjust the horizontal space as needed
        \begin{subfigure}{\imagewidth}
            \centering
            \includegraphics[width=\linewidth]{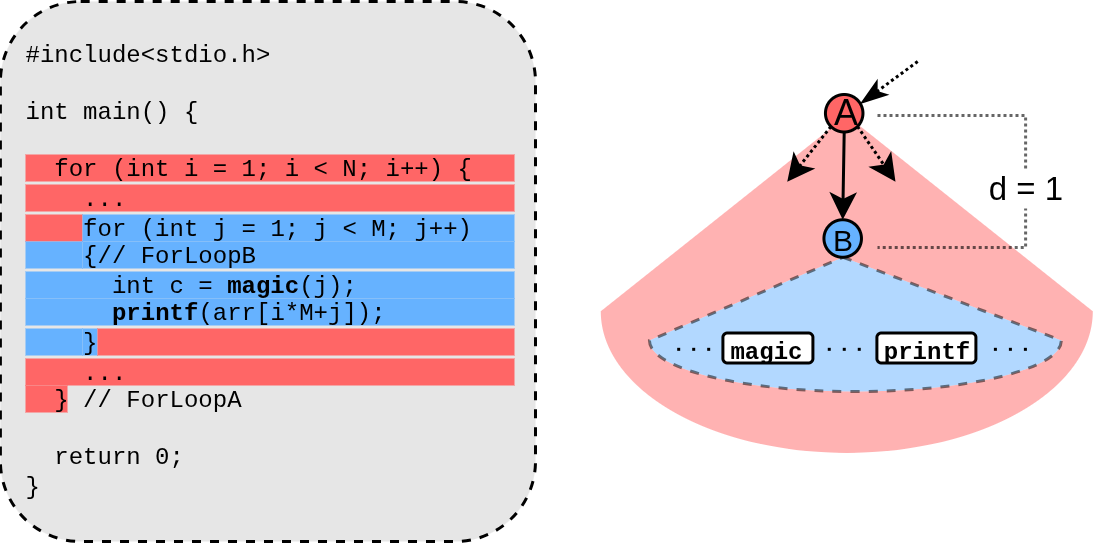}
            \caption{{\sffamily\textbf{Nesting}} matches ancestor \texttt{ForLoopA} and child \texttt{ForLoopB}. \texttt{ForLoopA} is also the context.
            }
            \label{fig:nesting-example}
        \end{subfigure}
    \end{minipage}
    
    \begin{minipage}{\linewidth}
        \centering
        \begin{subfigure}{\imagewidth}
            \centering
            \includegraphics[width=\linewidth]{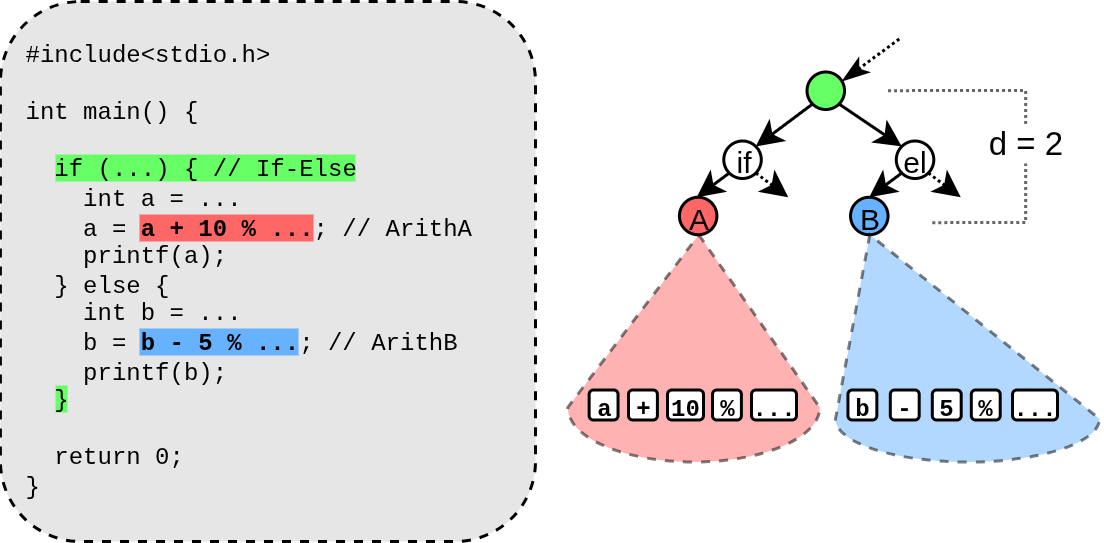}
            \caption{
            {\sffamily\textbf{Balanced}} matches the \textsc{If-Else} block \texttt{if(...)\{..\}\{..\}} as ancestor and \texttt{{a+10\%...}}, \texttt{b-5\%...} as descendants.
            } 
            \label{fig:balanced-example}
        \end{subfigure}
        \hspace{2em} % Adjust the horizontal space as needed
        \begin{subfigure}{\imagewidth}
            \centering
            \includegraphics[width=\linewidth]{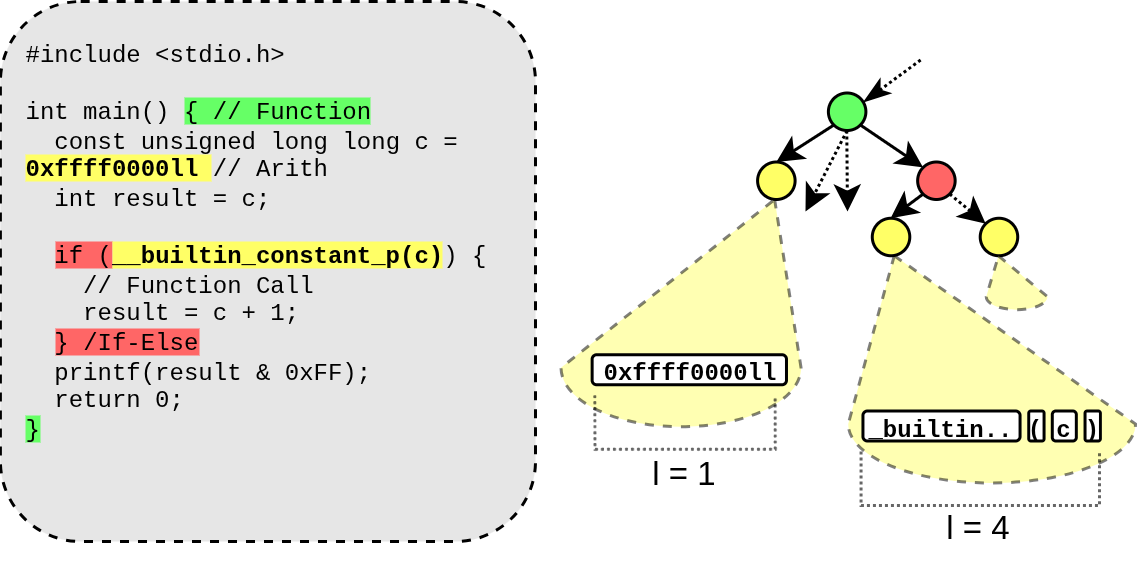}
            \caption{{\sffamily\textbf{Exists}} has multiple matches. We selectively highlight \texttt{0xff..} and \texttt{\_builtin..(c)}. Contextes are \texttt{Function} and \texttt{If-Else}.}
            \label{fig:exists-example}
        \end{subfigure}
    \end{minipage}
    \vspace{-.5em}
    \caption{Examples of composition styles matching on C construct trees.}
    \vspace{-1em}
    \label{fig:patterns-example}
\end{figure*}
\paragraph{Intuition.}
Many compiler optimizations fire only when specific \emph{structural} relationships hold among program constructs (e.g., a loop nested in a loop, or sibling expressions sharing an ancestor). Our fuzzer targets such relationships directly by elevating them to first-class \emph{composition styles}. A composition style specifies \emph{which types of constructs must co-occur, how they relate structurally, under what scope}, and \emph{which mutations (Section~\ref{sec:mutator}) can re-create} that structure.

\paragraph{Definition (Composition Style).}
Let $G$ be a construct tree and $N$ its constructs. Denote the \emph{type} of construct $n$ as $T(n)$. A composition style is
\[
C = \big(R_C,\; \langle \mathcal{T}_1,\dots,\mathcal{T}_k\rangle,\; \mathcal{T}_{\text{ctx}},\; p\big),
\]
\begin{enumerate}[label=(\roman*), leftmargin=2.1em]
\item $R_C \subseteq N^k$ is the $(k)$-ary structural relation targeted by $C$, relating $k$ constructs $\langle n_1, \dots, n_k \rangle$. For example, we define \textsf{Cousins} to be a binary relation between two constructs that share a common ancestor.
% Each style's relation is programmatically defined by a subroutine \texttt{C::scan(n)} that performs tree-traversal on \texttt{n} to identify constructs $\langle n_i \rangle$ that satisfy $R_C$
% (See \textit{Operationalization}) 

\item $\mathcal{T}_i$ is the set of admissible types (e.g., \textsc{Loop}, \textsc{FuncCall}) of the $i$-th construct, i.e., each matched construct $n_i$'s type must satisfy $T(n_i) \in \mathcal{T}_i$. This constrains the matching of a structural relation to certain types known for exposing optimization opportunity. Some composition styles impose additional constraints on matched types: for instance, \textsf{Cousins} requires that all matched constructs share the same type: $T(n_1) = T(n_2)$; while \textsf{Precedes} allows matched constructs of different types: $T(n_1) \neq T(n_2)$ is permitted.

\item $\mathcal{T}_{\text{ctx}}$ is the set of admissible types (e.g., \textsc{If-Else}, \textsc{Func}) of the context $n_{\text{ctx}}$, i.e., it must satisfy $T(n_\text{ctx}) \in \mathcal{T}_{\text{ctx}}$. 
The context $n_{\text{ctx}}$ is the lowest common ancestor construct of all the matched constructs $\langle n_i \rangle$ and must "contain" them (defined in Section~\ref{sec:mutator}). It defines the \emph{scope} of the structural relationship. For example, \textsf{Cousins} requires the context to be \textsc{If-Else}, \textsc{Loop}, or \textsc{Func}, ensuring that the matched cousins co-occur within these meaningful program structures rather than at arbitrary locations in the construct tree.
\item $p$ is a set of additional predicates on $n_1, \dots, n_k$ that filter matches and constrain the structural relationship. These predicates are essential because the base structural relations are often too broad, leading to trivial or uninteresting matches. For example, without predicates, \textsf{Cousins} would match \emph{any} two loops in a function since they technically share the function body as a common ancestor. To address this, each composition style defines specific predicates: \textsf{Cousins} bounds the generational distance between constructs to ensure meaningful proximity, while \textsf{Balanced} limits the branch depth explored under an \textsc{If-Else} construct.
% \item $M$ is a set of supported mutations/crossovers. Given a \emph{match} $m$ in tree $G$~\jw{what is a \emph{match}? Should be defined.} \ez{There is no way for the definitions to be not cyclic... match is defined below}, a supported mutation 
% $\mu\in M$ is picked and instantiates an edit generator $\mu_m(P) \Rightarrow P'$ that \emph{reconstructs the same style} in a target program $P$.
% Details appear in Section~\ref{sec:mutator}.
\end{enumerate}

%\jw{It's fine to throw out the symbol, but you need to give an example to explain what each symbol exactly means. I'm lost. What is }
% \hj{I might need a few more passes of the paper to appreciate it, but right now, I find this formalism confusing and I think it raises more questions than answers. We introduced new symbols and used new terms, e.g. "meta-heuristic mutation" and "construct types", that is never referenced again in the paper, or referenced with a different words, e.g., "edit generator" vs "mutator". It may be also be good to allude to  where each piece of information would come from and would be used, e.g., the custom program constructs annotated by the user, the program transformations already pre-configured for each style. }

% \ez{I removed unnecessary terms and added motivations for each symbol to help the readers understand why we need them.}

% \ez{I also added some running examples to help understand the definitions.}

% \hj{We might also want to provide a precise deinfition of a "construct type", e.g., say that its an AST node (or was it a parse tree node? I don't remember clearly), or a custom construct from Section 3.1}

% \ez{I unified the terms. "program construct" and "construct" mean the same thing. A parse tree node is of a certain "construct" such as \textsc{Loop}, \textsc{FuncCall}, \textsc{Arithmetic}, etc. I removed "construct type".}

\paragraph{Operationalization.}
\tool{} implements each style $C$ with a scanning routine that programmatically defines its structural relation $R_C$:
\[
\texttt{C::scan}(n) \rightarrow m; \quad\quad
m : \langle C, \langle n_1,\dots,n_k \rangle,n_{\text{ctx}}, p\rangle
\]
that traverses the subtree at root node $n$ and returns a \emph{match} $m$—a witness that the style $C$ holds at constructs $\langle n_i \rangle$ under context $n_{\text{ctx}}$ with predicates $p$ satisfied. Concretely, the returned tuple satisfies the style’s relation $R_C$, the type constraints $\langle \mathcal{T}_i\rangle$ and $\mathcal{T}_{\text{ctx}}$, and the predicates $p$. For each match, the fuzzer gathers the minimal $n_{\text{ctx}}$ that encloses the declarations and uses needed for scope well-formedness. If no such tuple exists inside subtree $n$, \texttt{C::scan} returns no result; In Phase 2~(Figure~\ref{fig:flowchart}), \tool{} runs \texttt{C::scan} on all construct trees from the optimization corpus to extract matches. 

\textbf{Instantiating composition styles to extract optimization opportunities.}
We extracted from \textit{seminal works} on optimizing transformations \cite{allen1971catalogue, Aho1970formal2optims} and \textit{compiler documentation}~\cite{llvmPasses, mlirPasses} a set of recurring structural preconditions that commonly trigger optimizations. 
Then, \textit{we instantiated six concrete composition styles} with fully specified relations, types, and predicates. Table~\ref{tab:patterns} shows the registry of these concrete styles. For brevity, we highlight four representative ones and illustrate how they capture common optimization-triggering conditions.

\paragraph{\textsf{Cousins}.}
Matches two constructs of the same type—\textsc{Loop}, \textsc{FuncCall}, \textsc{Arithmetic}, or \textsc{Logical}—that share the same ancestor. Optional parameters bound proximity: maximum generation distance $k$ and maximum token-distance $d$. Figure~\ref{fig:cousins-example} shows \texttt{ForLoopA}/\texttt{ForLoopB} with $k=0$, $d=0$. Captures redundancy from loop-level (\textbf{Loop Fusion} with the paired style in Fig.~\ref{fig:loopfusion-example-listing}) to instruction-level peepholes (\textbf{InstCombine}: adjacent add/sub, shifts, casts, comparisons).

\paragraph{\textsf{Nesting}.}
Matches same-type ancestry among control constructs—nested \textsc{Loop} and \textsc{If-Else}. Optional parameter \(d\) bounds nesting depth (tree distance between the matched nodes). Figure~\ref{fig:nesting-example} illustrates a match. Targets control-flow–simplifying optimizations such as \textbf{Loop Flatten}, \textbf{Loop Interchange}, \textbf{Loop Unroll and Jam}, and \textbf{Constraint Elimination}.

\paragraph{\textsf{Balanced}.}
Matches an ancestor \textsc{If-Else} with two descendant constructs of the \emph{same} type—\textsc{Loop}, \textsc{FuncCall}, \textsc{Arithmetic}, or \textsc{MemRef}—appearing in different branches. Optional parameter $d$: maximum branch depth explored under the \textsc{If-Else}. Figure~\ref{fig:balanced-example} shows similar arithmetic in two branches, making them potential candidates for hoisting. Captures replicated work across guarded paths; common targets include \textbf{GVN} (hoisting/sinking equivalent computations) and \textbf{Jump Threading} (simplifying conditionals around calls).

\paragraph{\textsf{Exists}.}
Matches a \emph{single} construct drawn from Table~\ref{tab:patterns} of type: \textsc{Func}, \textsc{Loop}, \textsc{If-Else}, \textsc{FuncCall}, \textsc{MemRef}, \textsc{Vector}, \textsc{Arithmetic}. Minimum lexer nodes $l$ limits the match to nodes of at least $l$ tokens. 
Useful to constrain fuzzing scope (e.g., scalar vs.\ loop) and surface \emph{magic numbers}. Certain rare-typed constants are known to trigger issues in \textbf{Constant Propagation} and \textbf{Constant Hoisting}. Figure~\ref{fig:exists-example} highlights two matches (\textsc{Arithmetic}, \textsc{FuncCall}).

\subsection{Mutators}\label{sec:mutator}

Having extracted composition styles (matches) from donor programs, \tool{} synthesizes mutators to apply on programs from the seed corpus (Phase 3). By reconstructing composition styles within different program contexts, \tool{} aims to trigger \textit{variations} of optimization logic.
\tool{} implements 4 high-level program transformations: 2 mutations that rebuild structural relations using only recipient material, and 2 crossovers that transplant donor fragments into the recipient.

\textbf{Mutator.} A mutator $\mu$ is a generator with signature
$\mu_m(P)\Rightarrow P'$, where $m=\langle C, \langle n_1,\dots,n_k \rangle, n_{\text{ctx}}, p\rangle$
is a match extracted from a donor construct tree in the optimization corpus, and $P$ is a \emph{recipient} construct tree in the seed corpus.

Given a match $m$ and a recipient $P$, a mutator $\mu$ performs these steps:

\smallskip
\noindent\emph{\textbf{\circled{R} Replicate \& \circled{M} Move} — Mutate \underline{within} the recipient (no donor construct inserted).}

\begin{enumerate}[leftmargin=1.6em,itemsep=2pt]
\item \textbf{Partialization.} Remove one matched construct
$n_i \in \langle n_1,\dots,n_k \rangle$ from context $n_{\text{ctx}}$ to obtain a \emph{partial context}
$n'_{\text{ctx}}$ and an \emph{anchor set} $A= \langle n_1,\dots,n_k \rangle \setminus \{n_i\}$.

\item \textbf{Context match.} i) Find candidate context constructs in $P$ with same type as partial context $n'_{\text{ctx}}$ and rank by structural similarity to $n'_{\text{ctx}}$
(KLR similarity~\cite{synthfuzz}), then pick the most similar recipient context $n^{P}_{\text{ctx}}$. ii)
structurally match anchor constructs to constructs in the recipient context, $A \mapsto A^{P}$, that is, $(n_a \mapsto n^P_a, \dots)$. iii) Identify a
\emph{location} $\ell$ within $n^{P}_{\text{ctx}}$ corresponding to the missing construct $n_i$.

\item \textbf{Transform (this distinguishes the 2 mutations).} \begin{description}[leftmargin=1.6em,itemsep=2pt]
\item[\circled{R} Replicate.] With anchors $A^{P}$ and location $\ell$,
choose some $a^{P}\in A^{P}$ and \emph{clone} it into $\ell$ to re-create the missing counterpart. Applicable when composition style requires that all matched constructs share the same type (Table~\ref{tab:patterns}:\,\textsf{Cousins}; \textsf{Nesting}; \textsf{Sequence}; descendants in \textsf{Balanced}).

\item[\circled{M} Move.] Find all constructs $x$ in the \emph{entire} recipient tree with $T(x)=T(n_i)$; rank candidates by lexical edit distance to the removed $n_i$; \emph{move} the best-scoring $x$ to $\ell$. Applicable for both same-type styles and "heterogeneous" styles that don't require matched constructs to have the same type (Table~\ref{tab:patterns}:\,\textsf{Precedes}).
\end{description}
\end{enumerate}

\smallskip
\noindent\emph{\textbf{\circled{I} Insert \& \circled{P} Replace} — Crossover (transplant) donor constructs into the recipient}
\begin{enumerate}[leftmargin=1.6em,itemsep=2pt]
\item \textbf{Context match.} Same as Step 2 above, except that no partialization is needed; match the full context $n_{\text{ctx}}$ and all constructs $\langle n_1,\dots,n_k \rangle$ to $n^{P}_{\text{ctx}}$ and $A^{P} = \langle n^P_1,\dots,n^P_k \rangle$.

\item \textbf{Transform (this distinguishes the 2 crossovers).} 
\begin{description}[leftmargin=1.6em,itemsep=2pt]
\item[\circled{I} Insert.] After matching $n^{P}_{\text{ctx}}$ and $A^P$, \emph{insert} each donor construct $n_i$ next to its match $n^P_i$.

\item[\circled{P} Replace.] After matching $n^{P}_{\text{ctx}}$ and $A^P$, \emph{replace} each $n^P_i$ with donor construct $n_i$.
\end{description}
\end{enumerate}

\paragraph{Parameterized mutation.}
Edits—especially crossovers—can easily violate scope or typing (e.g., transplanting an expression $a+b$ into a context that lacks declarations for $a,b$). 
We therefore perform parameterized mutation~\cite{synthfuzz}: every edit is re-parameterized against the recipient context so that free uses are bound to in-scope declarations and types are compatible. That is, the contexts we extract during style scanning (Section~\ref{sec:pattern}) and during context matching (Step~2 in Section~\ref{sec:mutator}) are required to satisfy a declaration–use \emph{containment} property.
We define that a context $n_{\text{ctx}}$ contains constructs $\langle n_1,\dots,n_k\rangle$, if all \emph{uses} appearing in the constructs are resolvable to \emph{declarations} within $n_{\text{ctx}}$ and types must be compatible.
This is enabled by grammar annotations introduced in Section~\ref{sec:constructs}. Special constructs \texttt{USES\_,DEFS\_,TYPES\_} must be defined for \tool{} to identify declaration-use relations and support parametrized mutation.

When cloning an anchor $n^{p}\!\in\!A^{P}$ into $\ell$ (Replicate) or moving a construct $x$ into $\ell$ (Move), the mutator rewrites the construct's \texttt{USES\_} to names drawn from \texttt{DEFS\_} that are visible in $n^{P}_{\text{ctx}}$ and type-compatible per \texttt{TYPES\_}.
When inserting or replacing with donor constructs $n_i$, the mutator re-parameterizes the donor’s \texttt{USES\_} to bind to \texttt{DEFS\_} visible in $n^{P}_{\text{ctx}}$ (and checks \texttt{TYPES\_}). If no compatible binding exists, the edit is rejected. Parameterized mutation substantially increases edit validity rates while keeping edits local to the matched context.

\begin{figure}[tb]
    \centering
    \includegraphics[width=0.97\linewidth]{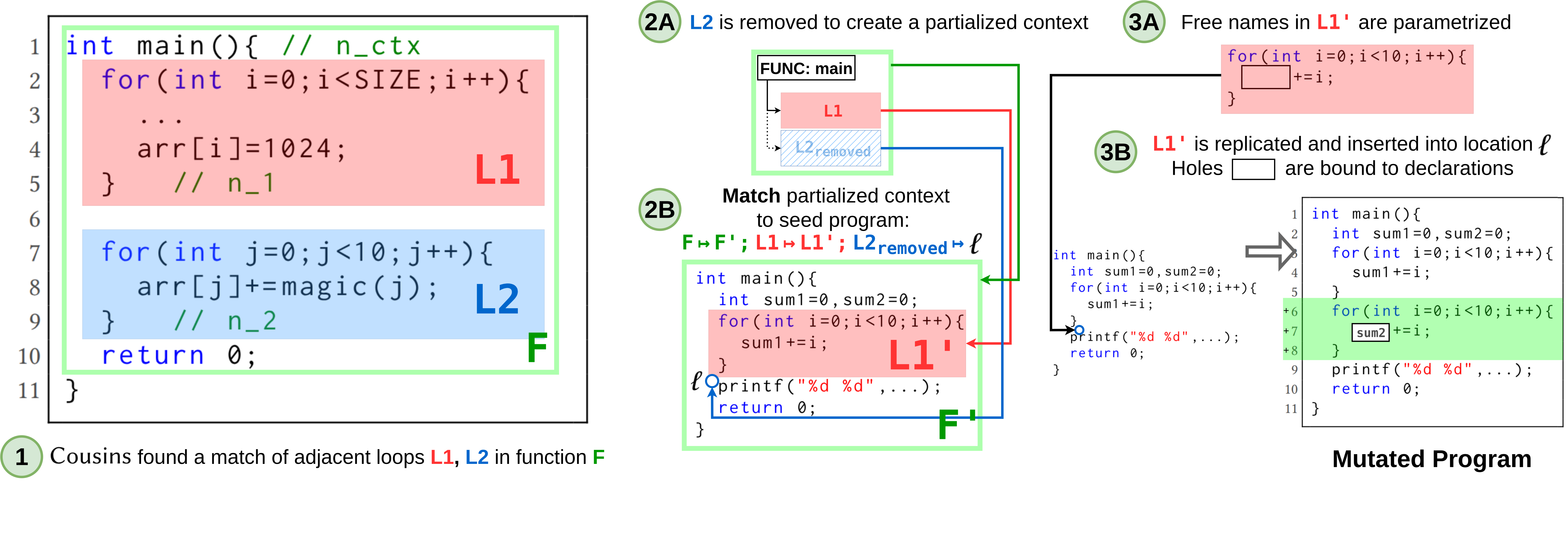}
    \caption{A working example of \tool{}. First, the \textsf{Cousins} style matches two adjacent loops $L_1, L_2$ within function context $F$ in the donor; Partialization removes $L_{2}$ to create a partialized context. Then, the partialized context and its constructs are mapped to structurally similar constructs in the seed program: $F\mapsto F',L_1\mapsto L_1',L_2{\text{(removed)}}\mapsto \ell$. Next, \textbf{Replicate} clones the matched segment $L'_{1}$, inserts it at location $\ell$, and substitutes the free variable name \textbf{sum1} with \textbf{sum2}.}
    \label{fig:loop-fusion-mutation-example}
\end{figure}

\paragraph{Example.}
Figure~\ref{fig:loop-fusion-mutation-example} illustrates an end-to-end application of \circled{R}\,\textsc{Replicate} for the \textsf{Cousins} style. In the donor (left), the scanner identifies a match \(m=\langle \textsf{Cousins}, \langle L_1,L_2\rangle, F,\; p\{k=0,d=0\}\rangle\): two same-type loops \(L_1\) (red) and \(L_2\) (blue) enclosed by the function context \(F\) (green). For simplicity, suppose the names \texttt{SIZE,arr,magic} are all declared in the function scope, i.e. $F$ "contains" the constructs $L_1,L_2$. \emph{Partialization} removes \(L_2\), retaining \(L_1\) as the anchor and yielding a hole for the missing cousin. In the recipient (middle), context matching selects the most similar function \(F'\), locates the anchor loop \(L_1'\), and determines a position \(\ell\) corresponding to the removed cousin (\(F\mapsto F'\), \(L_1\mapsto L_1'\), \(L_2\mapsto \ell\)). \textsc{Replicate} then clones \(L_1'\) into location $\ell$, producing the mutated program and, via parameterized mutation, rebinds free \texttt{USES\_} of aggregator \texttt{sum1} within \(L_1'\) to in-scope \texttt{DEFS\_}, producing a second loop that updates \texttt{sum2}. Note that \texttt{i} is \texttt{USES\_} but it is declared within the loop construct itself, so there is no need to rebind it. The transformation reconstructs the \textsf{Cousins} relation in the recipient and exposes a canonical loop-fusion opportunity.

\section{Evaluation}
In our study, we examine the following research questions:
\begin{enumerate}[label=\textbf{RQ\arabic*:}]
    \item \label{rq:coverage} How effective is \tool{} at testing optimizations compared to other fuzzers, in terms of coverage?
    % How does \tool{} generalize to different categories of optimizations?
    \item \label{rq:mode_comparison} How does \tool{}'s targeted fuzzing mode compare to whole pipeline fuzzing mode?
    \item \label{rq:trigger} What are \tool{}'s strengths \& limitations at triggering optimizations of different characteristics?
    \item \label{rq:ablation} What are each composition style and mutator's bug finding capability?
    \item \label{rq:bug-findings} \tool's bug findings.
\end{enumerate}

We implemented \tool{} on top of Grammarinator in \loc\, lines of code, including \locCAnnot\, and \locMlirAnnot\, lines of definitions of C (Figure~\ref{lst:gram_annot}) and MLIR language constructs, respectively. Annotating grammar is a one-time effort, and we estimate that it would take around two hours, for someone experienced with grammar-based fuzzing to define the program constructs for a new language.

\subsection{Experiment Design}

\subsubsection{Compilers and optimizations under test}
We evaluate \tool{} on 2 optimizing compilers, namely, LLVM and MLIR. LLVM is a production-quality optimizing compiler for languages such as C/C++. It contains the most compiler optimizations, and active optimization research is usually first committed there. MLIR is a modular compiler framework that is gaining popularity as a platform for deep learning compilers. Unlike LLVM which uses a single, monolithic intermediate representation (IR), MLIR allows developers to extend its IR with custom operations, types, and optimizations, which are logically grouped into \textit{dialects} such as \texttt{vector,affine,gpu}. MLIR's optimizations are mostly dialect-specific. 
We picked MLIR to demonstrate that \tool{}'s approach generalizes beyond high-level imperative programming languages like C. 

As explained in Section~\ref{sec:approach}, \tool{} supports both fuzzing entire optimization pipelines and targeted fuzzing of individual optimizations. We evaluate \tool{} on both modes for LLVM and only targeted fuzzing of individual optimizations for MLIR, because MLIR core dialect optimizer doesn't quite have an encompassing pipeline. For whole pipeline fuzzing of LLVM, we run \texttt{clang -O3} on \tool{}'s mutated programs for 24 hours. Using this traditional evaluation setting \cite{mendoza2023grayc, li2024creal} ensures a fair comparison.
For targeted fuzzing of individual optimizations, we run each optimization pass for one hour in RQ1 and four hours in RQ2. The shorter fuzzing budget is because individual optimization is small (mostly <10k LOC).

To "target-fuzz" individual optimizations, we need to perform sampling since technically, LLVM and MLIR each implements over 200 optimization passes. For each sampled optimization, we craft a mini "preparation" pipeline, e.g., run \texttt{-loop-simplify,-loop-rotate} before loop optimizations. We sampled \llvmnumoptim\ (LLVM) and \mlirnumoptim\,(MLIR) optimizations. We refer to them as the \emph{sampled optimizations}.  
Our selection covers 4 LLVM optimization categories (Scalar, Loop, Instcombine, Interprocedural) and \mlirnumdialects\, MLIR core dialects: \texttt{affine}, \texttt{vector}, \texttt{arith}, \texttt{scf}, \texttt{async}, \texttt{gpu}, \texttt{memref}.
We believe the diversity is enough and comparable to the literature on compiler optimization testing \cite{yang2024whitefox}.

\subsubsection{Optimization and Seed corpus} From the sampled optimizations, we distill the source code of each optimization pass into metadata, function declaration, and docstrings. Then we prompt GPT4o-mini to collect 100 valid programs per optimization pass as the its own optimization corpus. 
We collect two seed corpus of 1059 C programs and 3601 MLIR programs from the literature and compiler test suites \cite{mendoza2023grayc, yang2011csmith, synthfuzz}. In targeted fuzzing mode, we use each pass's own optimization corpus (100). In pipeline fuzzing mode, we pool all sampled optimizations' corpus (3700).

\subsubsection{Evaluation Metrics}
We use the following metrics:
\begin{itemize}    
    \item \textbf{Branch Coverage} We built the optimizing compilers (\texttt{clang}, \texttt{opt} \& \texttt{mlir-opt}) with coverage instrumentation \cite{llvminstrument}.
    \item \textbf{Optimization Trigger Throughput} is the total number successful program transformations triggered/applied in four hours. This is a standard evaluation metric in the domain of fuzzing compiler optimizations \cite{yang2024whitefox, zhou2024polyjuice, xie2025mopfuzzer}.
    For example, LoopFuse's \#transforms is \#loops fused, and GVNHoist's \#transforms is \#instructions hoisted or removed.
\end{itemize}

\subsubsection{Baselines}

We evaluate \tool{} against Grammarinator \cite{Hodovan2018grammarinator}, SynthFuzz \cite{synthfuzz}, GrayC \cite{mendoza2023grayc}, and MLIRSmith \cite{Wang2023mlirsmith}. To ensure fairness, we \textit{combine} the optimization and seed corpus to use as the fuzzing corpus of Grammarinator, SynthFuzz, and GrayC. Grammarinator represents a baseline grammar-based fuzzer. 
SynthFuzz is a grammar-based fuzzer that synthesizes custom mutations that preserve semantic validity. We pick SynthFuzz because it is 1) novel in synthesis of grammar mutation, and 2) shown to be the state-of-the-art MLIR fuzzer \cite{synthfuzz}. GrayC is a coverage-guided, mutational fuzzer targeting the C language. It implements a set of semantic-aware, custom mutators, such as replacing an arithmetic operator with another. We pick GrayC because it's the state-of-the-art C Compiler fuzzer \cite{mendoza2023grayc}. We also compare against GrayC's variant without coverage guidance in RQ1 as it uses the coverage as metric. MLIRSmith is a custom, generator-based fuzzer that targets MLIR's core dialects. It requires developers to manually implement custom generator for each dialects. We pick MLIRSmith as a baseline for generational MLIR fuzzers.

\subsection{RQ1: Is \tool{} effective at testing optimizations compared to other fuzzers?} 
\label{sec:rq1}
\begin{figure*}[tb]
    \begin{minipage}[t]{0.24\textwidth}
        \includegraphics[trim={0.3cm 1cm 0 0.5cm},clip,width=0.98\linewidth]{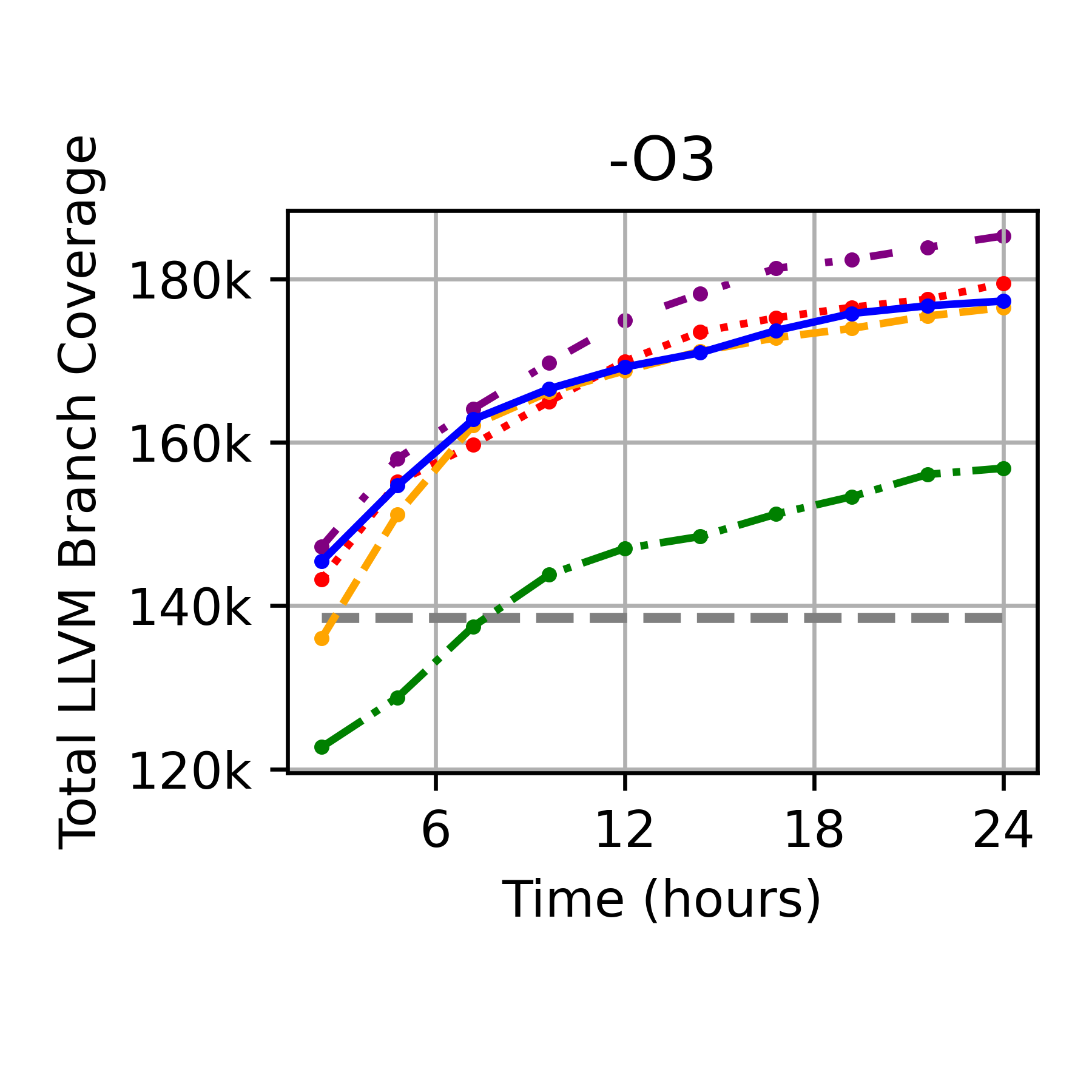}
        \subcaption{Whole pipeline fuzzing}
        \label{fig:llvm-coverage-o3}
    \end{minipage}
    \hspace{-0.03\textwidth}
    \begin{minipage}[t]{0.74\textwidth}
        \centering      
        \includegraphics[trim={3.5cm 0 4.5cm 0},clip,width=0.98\linewidth]{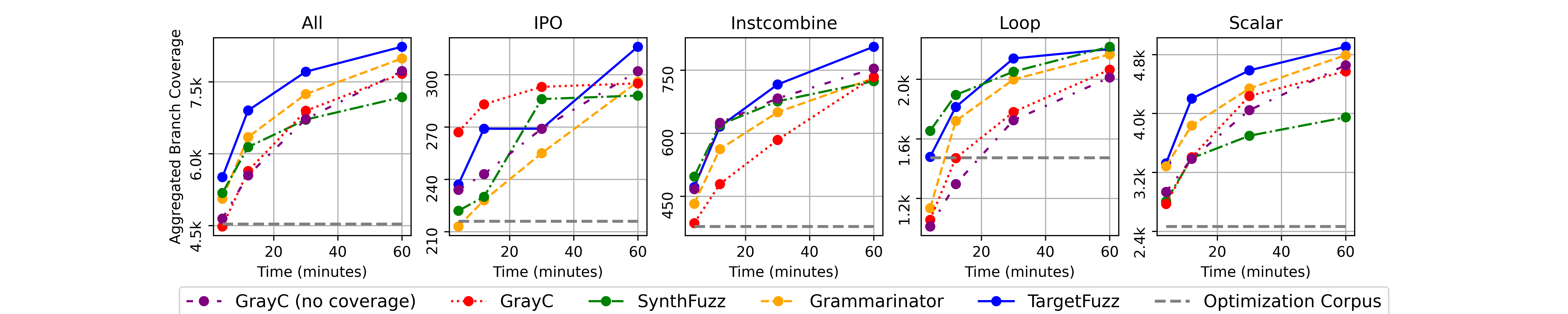}
        \subcaption{Targeted optimization fuzzing}
        \label{fig:llvm-coverage-targeted}
    \end{minipage}
    
    \caption{\textbf{RQ1}: (a) Coverage on entire LLVM codebase when fuzzing with -O3 pipeline for 24 hours. (b) Coverage on targeted optimizations, aggregated across 4 categories. Left to right: All optimizations, Interprocedural (IPO), InstCombine, Loop, and Scalar optimizations. In the whole pipeline fuzzing mode, \tool{} achieves lower coverage than GrayC(w/wo) coverage guidance when fuzzing the entire compiler; however, in the targeted fuzzing mode, \tool{} outperforms all baseline fuzzers on IPO, InstCombine, and Scalar families of optimizations. We show that the two fuzzing modes are complementary to each other in RQ2.}
    \label{fig:llvm-coverage}
    \vspace{-1em}
\end{figure*}

\begin{figure*}[tb]
    \centering
    % 0.80 and 0.17 linewidth for LLVM 4 types + MLIR all optimizations
    \includegraphics[trim={0cm 0 0cm 0},clip,width=\linewidth]{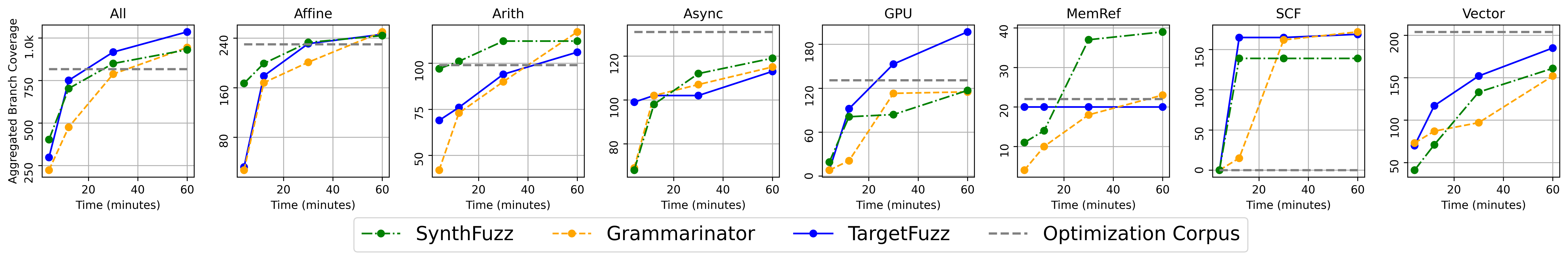}
        % \caption{LLVM Branch Coverage aggregated over four types of optimizations}
    \vspace{-1em} 
    \caption{\textbf{RQ1}: Coverage on targeted MLIR optimizations, summed across 7 categories. Left to right: All optimizations, Affine, Arith, Async, GPU, Memref, SCF, Vector. \tool{} achieves highest overall coverage and in dialects \texttt{gpu} and \texttt{vector}, while having comparable (<2\%) coverage in \texttt{affine}, \texttt{scf} as the best baseline.} 
    \label{fig:mlir-coverage}
\end{figure*}

For whole optimization pipeline fuzzing, we run \tool{} and baseline fuzzers for 24 hours on the \texttt{clang -O3} pipeline, then measure branch coverage of the entire compiler (directories under \texttt{clang/} and \texttt{llvm/}). For targeted fuzzing of individual optimizations, per optimization, we run each fuzzer for one hour and measure coverage only within that specific optimization's code.

On LLVM pipeline \texttt{-O3}, GrayC \footnote{GrayC wo. coverage achieves 3\% higher coverage than with coverage, consistent with GrayC's own evaluation results \cite{mendoza2023grayc}} fuzzer achieves the best overall coverage, followed by \tool{}, Grammarinator, then SynthFuzz, as shown in Figure~\ref{fig:llvm-coverage-o3}. 
This result aligns with expectations: GrayC is specifically designed for C/C++ programs with hand-coded mutations that operate on semantic-rich, typed ASTs to ensure high test validity. This explains why it is the most effective at testing C compiler as a whole. However, we argue that \tool{} is built with two distinctive goals: (1) unlike GrayC, \tool{} is a language-agnostic fuzzing framework that can be easily adapted to new languages (like MLIR), and (2) it is designed to be more effective at testing individual optimizations, which we will show in the rest of RQ1 and RQ2. 

Notably, \tool{} achieves the highest coverage among grammar-based fuzzers. Upon a closer look, \tool{} outperforms Grammarinator a lot at the beginning of the fuzzing campaign: at the 2.5 hour mark, \tool{} achieves 7\% higher coverage, but this advantage levels out in the long run. After 24 hours, \tool{} is only marginally better than Grammarinator. This is not surprising given the fact that Grammarinator's general mutations indeed \textit{subsume} \tool{}'s targeted, constrained mutators, that they eventually converge to similar coverage. However, \tool{} demonstrates an advantage in short fuzzing runs.

Under the targeted fuzzing setting, in LLVM, \tool{} achieves $3.0\%, 14.7\%, 7.3\%, 6.5\%, $ higher total coverage than Grammarinator, SynthFuzz, GrayC, and GrayC (without coverage), aggregating all sampled optimizations. 
To show generality, we further group the {\llvmnumoptim} LLVM optimizations into four categories: Scalar, Loop, Interprocedural (IPO), and InstCombine and report the aggregate coverage in Figure~\ref{fig:llvm-coverage-targeted}. \tool{} achieves the highest coverage on InstCombine, IPO, and Scalar optimizations. On Loop optimizations, SynthFuzz achieves $0.7\%$ higher coverage than \tool{}. 

In MLIR targeted fuzzing, \tool{} achieves $9.7\%$ and $11.2\%$ higher coverage than Grammarinator and SynthFuzz, respectively. We evaluated {\mlirnumoptim} optimizations across {\mlirnumdialects} MLIR dialects. As shown in Figure~\ref{fig:mlir-coverage}, \tool{} achieves the highest aggregated branch coverage across all optimizations. Specifically, \tool{} achieves best performance on the \texttt{gpu} and \texttt{vector} dialects and comparable coverage on the \texttt{affine} and \texttt{async} dialects (<margin of $2\%$).
MLIRSmith achieves near-zero coverage; we discuss the reasons for this in Section~\ref{sec:rq3}.

\begin{resultbox}
\tool{} achieves the highest overall coverage, in both LLVM and MLIR, under the targeted fuzzing setting, where it performs consistently better than other language-specific fuzzers, and grammar-based fuzzers. Under the whole pipeline fuzzing setting, \tool{} is inferior to GrayC (without coverage) and perform similarly to GrayC with coverage.
\end{resultbox}

\subsection{RQ2: How does \tool{}'s targeted fuzzing mode compare to whole pipeline fuzzing mode?} \label{sec:rq2}
\begin{figure}[t]
  \centering
  \begin{minipage}[t]{0.45\linewidth}
    \centering
    \includegraphics[width=\linewidth]{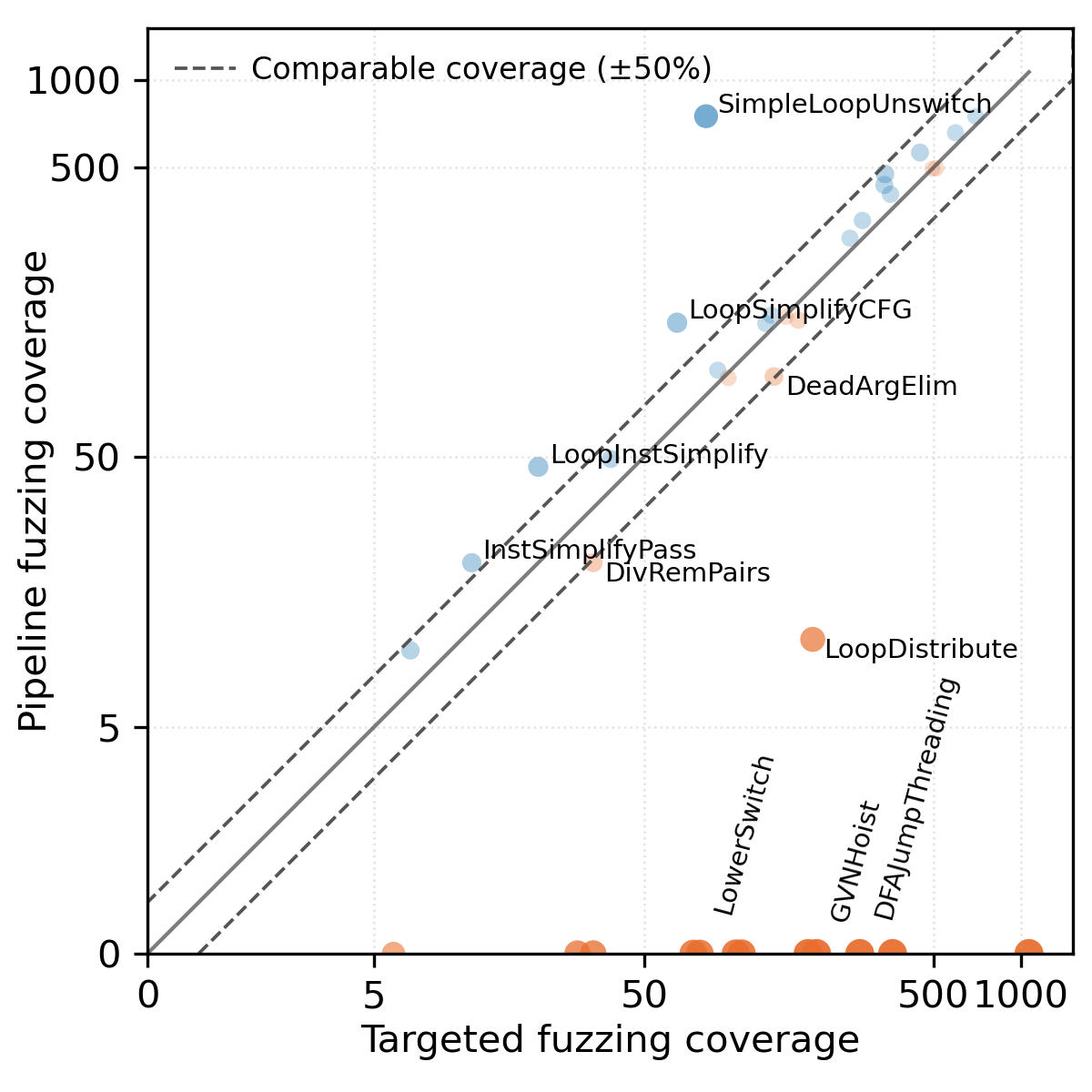}
    % \vspace{-1.3em}
    \captionof{figure}{\textbf{RQ2}: \tool{}'s coverage on sampled optimizations in targeted mode vs. pipeline mode}
    \label{fig:targeted_vs_pipeline}
  \end{minipage}\hfill
  \begin{minipage}[t]{0.45\linewidth}
    \centering
    \includegraphics[width=\linewidth]{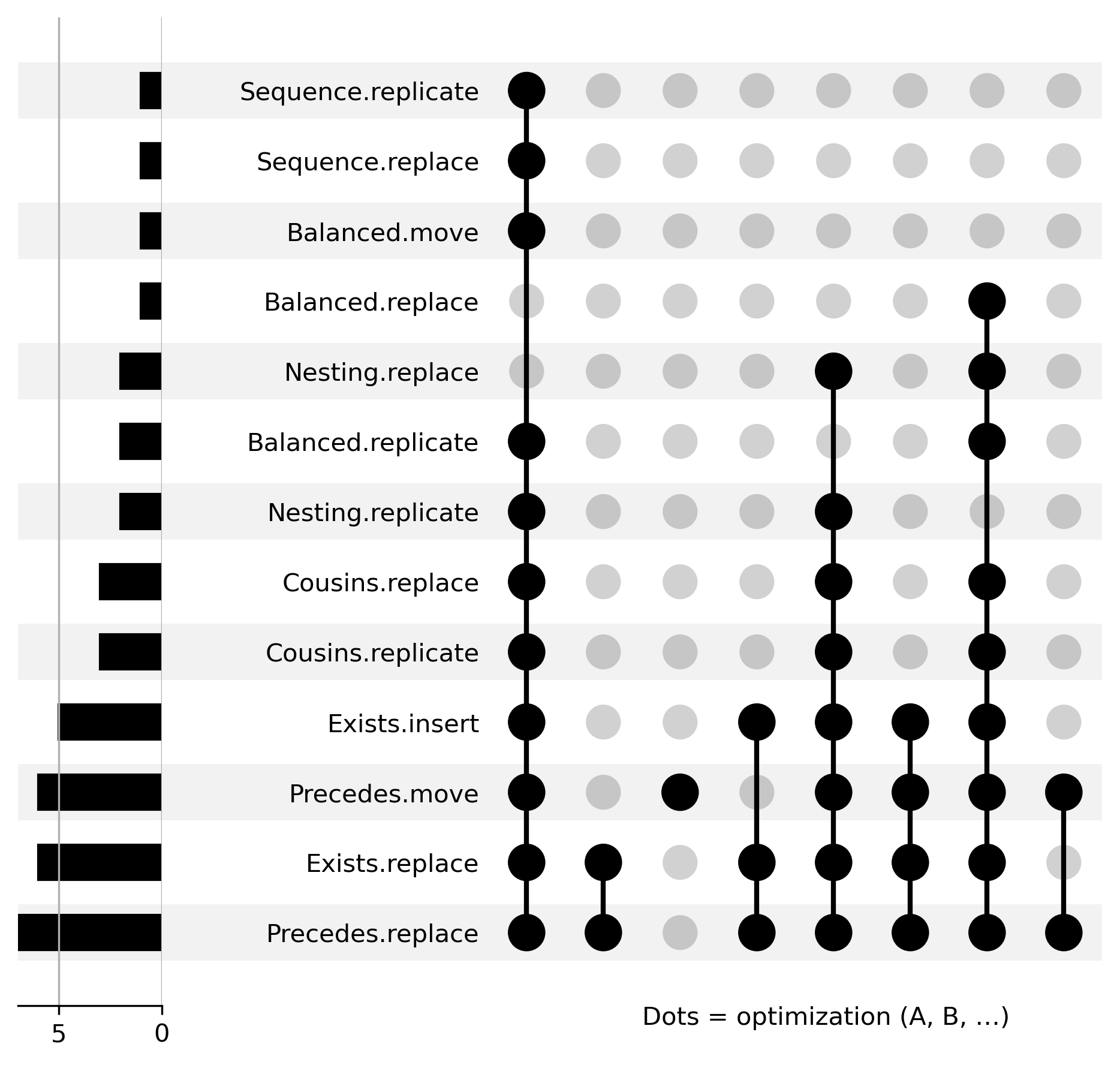}
    % \vspace{-1.3em}
    \captionof{figure}{\textbf{RQ4}: buggy optimizations flagged by each composition~\&~mutator.}
    \label{fig:venn_compmut}
  \end{minipage}
  % \vspace{-1.3em}
\end{figure}

In RQ1, we showed that under the whole pipeline fuzzing setting, \tool{} is comparable to GrayC, but inferior to GrayC without coverage, in terms of overall coverage of the LLVM compiler codebase. This is the coverage experiment that most compiler fuzzers carry out. But this result doesn't necessarily translate to effectiveness at testing optimizations. In particular, we would like to answer the question: \textit{Is there a need for targeted fuzzing of individual optimizations, that it complements what whole pipeline fuzzing cannot achieve?} 

First, we take a closer look at coverage of LLVM's optimization module under this mode. The optimization module under \texttt{llvm/lib/Transforms} contains 300 files, including all the transformation passes and their utilities. The best-performing baseline GrayCnoCov achieves 38596 branches, while \tool{} achieves 36377 branches, a 6.1\% difference. GrayCnoCov achieves higher coverage on 55 files while \tool{} achieves higher coverage on 36 files. However, \textit{152 of the 300 files have zero coverage} from both fuzzers. This supports our hypothesis that the whole pipeline fuzzing approach leaves a lot of optimization logic untested.

We further compare \tool{}'s coverage in the two fuzzing modes. It is important to note that these two modes use different test harness (\texttt{clang -O3} vs. individual pass) and subsequently different fuzzing budget (24 hours vs. one hour). Thus, the numbers are not directly comparable, but we can still draw insights from the comparison. We show the coverage of sampled optimizations in the two fuzzing modes in Figure~\ref{fig:targeted_vs_pipeline}. The band around the diagonal indicates that for most optimizations, the coverage in the two modes are comparable. Notably, there are 12 passes that are entirely untested in the whole pipeline mode, but all of them are covered in targeted fuzzing mode. 

\begin{resultbox}
Targeted fuzzing of individual optimizations is necessary to complement whole pipeline fuzzing, as it significantly increases the number of optimizations tested.
\end{resultbox}

\subsection{RQ3: What are \tool{}'s strengths and limitations at triggering optimizations of different characteristics?} \label{sec:rq3}

\begin{figure*}[t]
    \centering
    \includegraphics[width=0.95\linewidth]{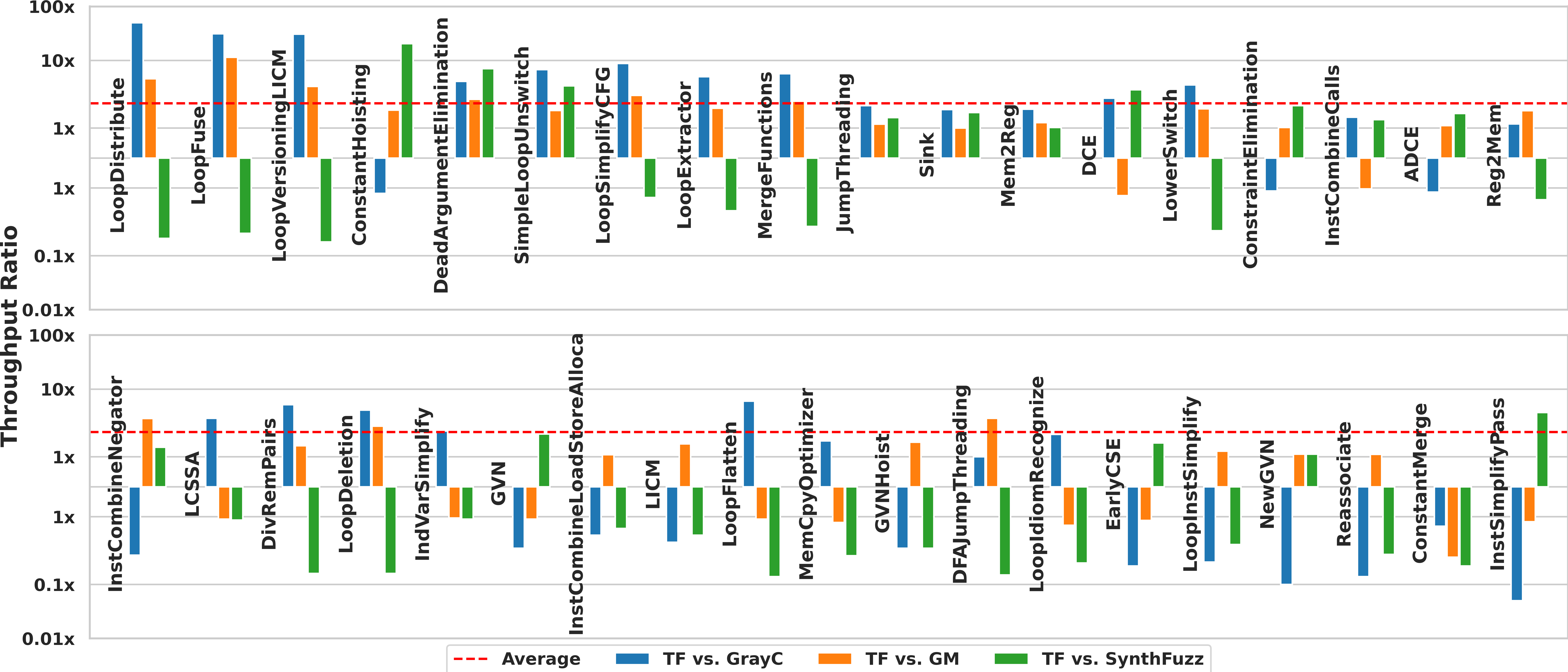}
    \caption{\textbf{RQ3}: \textit{LLVM} Ratio of Optimization Trigger Throughput. Each bar represents \tool{}'s throughput divided by baseline's throughput.
    Upward bar means \tool{} outperforms the baseline, and vice versa. The red line is the average ratio: \llvmtriggerimprv. On average, \tool{}'s trigger throughput outperforms GrayC, Grammarinator, and SynthFuzz by \llvmavgtriggerimprvgrayc, \llvmavgtriggerimprvgm, \llvmavgtriggerimprvsynth.}
    \label{fig:llvm_trigger}
    \vspace{0.5em}
    \includegraphics[width=0.95\linewidth]{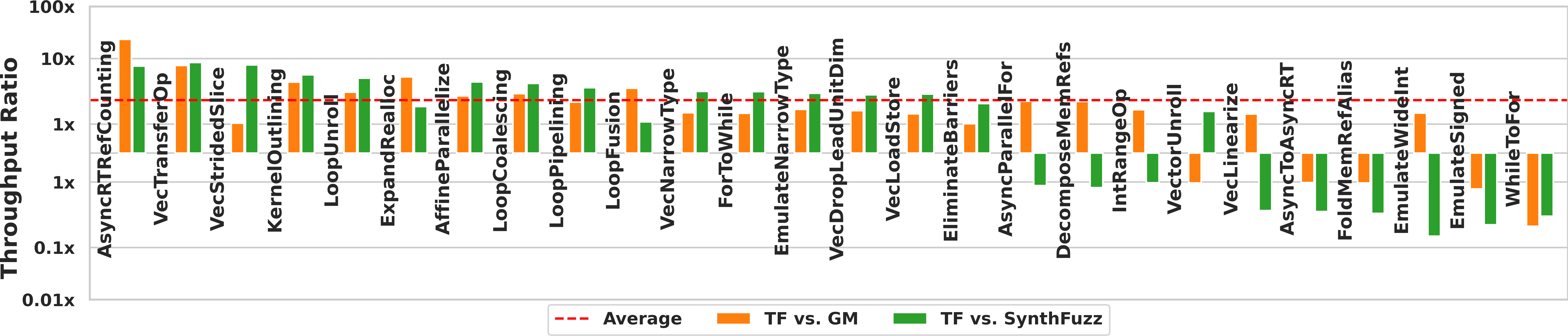}
    \vspace{-.4em}
    \caption{\textbf{RQ3}: \textit{MLIR} Ratio of Optimization Trigger Throughput. \tool{}'s outperforms GM and SynthFuzz by \mliravgtriggerimprvgm, \mliravgtriggerimprvsynth.}
    \vspace{-.6em}
    \label{fig:mlir_trigger}
\end{figure*}

To provide a more detailed assessment of \tool{}'s effectiveness in targeted fuzzing, we use the optimization trigger throughput metric. Widely used in prior work \cite{yang2024whitefox, zhou2024polyjuice, xie2025mopfuzzer}, it measures how frequently a fuzzer successfully triggers optimizations.
% We measured all fuzzers' trigger throughput in a 4 hr controlled experiment.
In Figure~\ref{fig:llvm_trigger} and Figure~\ref{fig:mlir_trigger}, we show \tool{}'s throughput improvements over the baseline fuzzers on all optimizations.

\tool{} outperforms Grammarinator by achieving higher trigger throughput on \llvmbettergm\, LLVM and \mlirbettergm\, MLIR optimizations, averaging \llvmavgtriggerimprvgm\, and \mliravgtriggerimprvgm\, throughput of Grammarinator. 
Because \tool{} is implemented on top of Grammarinator, these numbers directly reflect the gain of \tool{}'s composition-based, targeted mutations over naive grammar-based fuzzing, which is less effective at exercising deep optimization logic.

\tool{} achieves \llvmavgtriggerimprvsynth\, (LLVM) and \mliravgtriggerimprvsynth\, (MLIR) throughput of SynthFuzz. \tool{} outperforms SynthFuzz on \mlirbettersynth\, MLIR optimizations but SynthFuzz outperforms \tool{} on \llvmworsesynth\, LLVM optimizations.
SynthFuzz is a MLIR-inspired grammar-based fuzzer, leveraging parameterized mutation to transplant grammar subtrees to a different context~\cite{synthfuzz}. \tool{}'s mutations are more constrained to specifically target optimizations. In addition, program constructs allow \tool{} to identify declaration, usage, and types, so \tool{}'s parameterized mutation achieves higher validity (See Section~\ref{sec:mutator}). 
\tool{} produces C programs of \llvmtfvalidity\~validity compared to \llvmsynthvalidity~of SynthFuzz. 

MLIRSmith triggered zero MLIR optimizations, so we omit it in Figure~\ref{fig:mlir_trigger}. The problem with MLIRSmith's custom generator approach is twofold. MLIRSmith's programs have an extremely low validity rate of ~0.8\%. This is because MLIRSmith hardcoded the semantics of MLIR operations. MLIR's constant evolution easily renders such semantic specifications obsolete. Secondly, custom generators like MLIRSmith are built to maximize extensibility, i.e., being made versatile to target many language features. They are blackbox fuzzers that are unable to take advantage of domain knowledge (e.g. the optimization corpus) about the compiler under test, which has proven to be essential in testing deep compiler logic.

\tool{} outperforms GrayC on \llvmbettergrayc\, LLVM optimizations, averaging \llvmavgtriggerimprvgrayc\, throughput of GrayC.
However, \tool{}'s comparison against GrayC shows high variance: throughput ratio ranges from 50$\times$ to 0.05$\times$. Our analysis shows that \tool{} is better at triggering more constrained optimizations: for example, it outperforms GrayC on LoopFusion, DeadArgumentElimination, MergeFunctions, by 31.2$\times$, 4.1$\times$, and 5.6$\times$, respectively. 
GrayC performs better when the optimization is relatively "low-effort," i.e., has trivial triggering conditions. 
For example, GrayC outperforms \tool{} the most in InstSimplifyPass by $20\times$. It packages simple optimizations that remove redundant instructions, so any unoptimized C program have a high chance of triggering it. 
GrayC outperforms \tool{} by 10$\times$ on NewGVN (Global Value Numbering). 
This optimization removes equivalent expressions~\cite{newgvn}. 
GrayC happens to implement a \textit{Duplicate-Statement} mutator~\cite{mendoza2023grayc}, which duplicates a statement within the same block, easily triggering NewGVN. 

\begin{resultbox}
\tool{} is more effective than naive grammar-based fuzzer (Grammarinator), parametrized mutators (SynthFuzz) and language-specific fuzzer (GrayC) at triggering compiler optimizations. Compared to GrayC, \tool{} is particularly effective at triggering more constrained optimizations.
\end{resultbox}

\subsection{RQ4: What are each composition style and mutator's bug-finding capabilities?}
To evaluate the bug-finding capabilities of each composition style and mutator, we run \tool{} for 6 hours per sampled LLVM optimization, focusing on \emph{miscompilations}—the most severe optimization bugs. We use Alive2~\cite{lopes2021alive2} as the oracle to validate whether a triggered transformation is a miscompilation.
For each \{composition style, mutator\} pair, we report \emph{\# distinct optimizations} produced at least one Alive2-flagged miscompilation. Results appear in Figure~\ref{fig:venn_compmut}. In total, 8 optimizations are flagged for miscompilation. The most successful compositions are \textsf{Precedes} and \textsf{Exists}, and 5 optimizations are only discovered by them. It is expected that crossovers such as \textsf{Exists}+\textsf{Replace} and \textsf{Exists}+\textsf{Insert} are effective, as they are the most generic form of grammar-based mutations. However, \tool{}'s more constrained composition styles and mutations are also effective at exposing buggy optimizations. \textit{Notably, no single style-mutator pair alone flags all 8 buggy optimizations}, proving the necessity of multiple composition styles and mutators. 
\begin{resultbox}
\tool{}'s multiple styles and mutators are necessary to trigger a diverse set of optimization bugs.
\end{resultbox}

\subsection{Bug Findings}
\tool{} found {\numbugs} previously unknown bugs in LLVM and MLIR, shown in Table~\ref{tab:bug}. Among them, {\nummiddlebugs} are optimization bugs and {\numbackendbugs} are backend bugs; {\numwrongcode} are miscompilations and {\numcrash} are crashes. 
For reference, the optimization corpus itself only exposes one bug in LLVM, proving the effectiveness of \tool{}'s mutation.
We used two bug-finding oracles: 1) compiler crash (LLVM \& MLIR) and 2) Alive2 \cite{lopes2021alive2} (LLVM).
We also found 2 false-positive bugs in Alive2 and both have been fixed.

\begin{center}
  \begin{minipage}[t]{0.20\linewidth}\vspace{0pt}
    \centering\tiny
    \begin{tabular}{lrr}
      \toprule
      \textbf{Status} & \textbf{LLVM} & \textbf{MLIR} \\
      \midrule
      Reported      & 2 & 2 \\
      Confirmed     & 5 & 0 \\
      Fixed         & 2 & 5 \\
      NF & 2 & 0 \\
      \midrule
      Total         & 11 & 7 \\
      \bottomrule
    \end{tabular}
    \captionof{table}{Bug reports; NF: not-gonna-fix.}\label{tab:bug}
  \end{minipage}\hfill
  \begin{minipage}[t]{0.41\linewidth}\vspace{0pt}
    \centering
    \lstset{float=false, captionpos=b, aboveskip=0pt, belowskip=0pt}
\begin{lstlisting}[style=MLIRStyle,
      caption={\tool{} program crashes \textbf{VectorLinearize}},
      label={lst:vector_linearize_crash}]
func.func @extract_1d_constant()->(){
	%cst = arith.constant dense<[1,2]>:vector<2xi32>
	%0 = vector.extract %cst[0]: i32 fr vector<2xi32>
	func.return
}
\end{lstlisting}
  \end{minipage}\hfill
  \begin{minipage}[t]{0.27\linewidth}\vspace{0pt}
    \centering
    \lstset{float=false, captionpos=b, aboveskip=0pt, belowskip=0pt}
\begin{lstlisting}[style=CStyle,label={lst:loop-fusion-crash}, caption={\tool{} program crashes \textbf{Loop Fuse}}]
int main() {
	int sum1=0,sum2=0;
	for (int i=0;i<10;i++){
		sum1 += i;}
	for (int i=0;i<10;i++){
		sum2 += i*i;}
}
\end{lstlisting}
  \end{minipage}
  % \vspace{-.5em}
\end{center}

We present a \tool{}-found MLIR bug that was fixed. $\texttt{vector.extract}$ is an operation that extracts a $n-D$ vector at a $k-D$ position. \textbf{VectorLinearize} is an optimization pass that attempts to linearize all $n-D$ vectors to $1D$. 
\tool{} generated the program in Listing~\ref{lst:vector_linearize_crash} that crashed VectorLinearize, which incorrectly assumes that $k<n$ in all \texttt{vector.extract} operations, i.e., it must return vectors, and attempts to linearize the \texttt{vector.extract} on line 3 which in fact produces a scalar value of \texttt{i32}. \tool{} crafted this test program using the \textsf{Exists} style and \textbf{Insert} mutator. First, \tool{}'s identified that \textsc{Vector} constructs widely exists in the optimization corpus, matching a \texttt{vector.extract} operation. It then parameterizing its input argument. Upon matching the recipient context, the hole is concretized with the most structural similar node \texttt{\%cst}.
This mutation shows (1) how \tool{} uses optimization corpus to extract useful breadcrumbs (e.g., \texttt{vector.extract}) of the targeted optimizations, (2) parameterized mutation is critical in maintaining validty during crossover.

We present a \tool{}-found LLVM optimizer bug that was fixed. Optimization \textbf{LoopFuse} crashes on the program in Listing~\ref{lst:loop-fusion-crash}.
% when it is optimized with this pipeline: \texttt{mem2reg,loop-rotate,loop-fusion}. 
The two loops are fusable but \texttt{LoopFuser::performFusion} incorrectly asserts that the loops are illegal for fusion and crashes. This bug is both a crash and a missed optimization.

\section{Related Work }
While fuzzing is effective at exploring behaviors of code, 
a drawback is that it may only execute shallow code.
As such, it will struggle to test deep code, such as code related to optimization.

\noindent{\bf Grammar-based fuzzing.}
Grammar-based fuzzers, including Grammarinator~\cite{Hodovan2018grammarinator}, Nautilus~\cite{aschermann2019nautilus}, LangFuzz~\cite{holler2012langfuzz},
Skyfire~\cite{wang2017skyfire}, PolyGlot~\cite{chen2021polyglot}, and SynthFuzz~\cite{synthfuzz} take an input grammar to generate syntactically correct programs and mutate on parse trees. 
Generator-based fuzzers~\cite{padhye2019jqf} often rely on human effort to encode language-specific semantics. Even if semantically-valid inputs are generated, they do not necessarily satisfy the constraints for reaching deep code, such as optimizations. \tool{}, instead, takes advantage of composition styles, which allows it to attain higher coverage of compiler optimization code.

\noindent {\bf LLM-based fuzzing.} Fuzz4All~\cite{xia2024fuzz4all} was the seminal work to leverage LLMs to both generate and mutate tests programs for fuzzing compiler.  
MetaMut~\cite{ou2025metamut} automatically synthesizes semantics-aware mutators using LLMs through an elaborate, multi-stage prompting strategy.
\tool{} doesn't rely on LLMs for mutation. It only asks LLMs for an initial corpus of optimization tests, from where the mutations are derived through composition-styles and applied on a larger seed corpus at scale. This approach significantly reduces the number of slow and expensive LLM queries. In evaluation, \tool{}'s LLM-generated optimization corpus only exposes 1 bug; \tool{}'s bug findings come from its composition-style based mutations.

\noindent {\bf Semantic-aware fuzzing.}  
Some compiler fuzzers aim to more effectively test deep code by ensuring semantic well-formedness. MLIRSmith~\cite{Wang2023mlirsmith} encodes the semantics of MLIR dialects but cannot keep up with its constant evolution. YarpGen~\cite{Livinskii2020yarpgen} generates C/C++ programs that avoid undefined behaviors by static analysis and conservatively replacing unsafe operations. YarpGen v2~\cite{Livinskii2023yarpgenv2} added UB-free loop generation to target loop optimizations. GrayC~\cite{mendoza2023grayc} hand-coded a small set of mutators on semantic-rich, typed-AST to guarantee program well-formedness.
Ensuring semantic well-formedness is an orthogonal problem; \tool{} focuses on satisfying the structural constraints necessary for testing compiler optimizations. Our evaluation shows that structural constraints is easier to achieve, i.e., without the need for custom, semantic generators/mutators, and are more important than well-formesness in testing optimizations.

In theory, such constraints can be specified explicitly using ISLa~\cite{steinhofel2022isla}: a specification language for generating structured inputs. However, doing so one-at-a-time for each optimization, is prohibitively labor-intensive. \tool{} bypasses this problem by automatically extracting composition styles from the optimization corpus, essentially loosely-overapproximating these structural constraints.

The closest idea to \tool{} is Creal~\cite{li2024creal}, which injects functions of real-world code snippets into randomly generated programs (seed corpus) to increase test diversity. Creal essentially implements one "function injection" mutator: replace expressions in a seed program with calls to extracted real-world functions, but it ensures the 1) semantic equivalence and 2) well-formedness of the mutation, through a mix of profiling and static analysis. \tool{}'s program transformations are more general, e.g. function injection can simply be achieved by the \textsf{Exists} composition and the \textsf{Replace} mutator. However, \tool{} does not aim for semantic equivalence or well-formedness. 
\tool{} and Creal share a similar goal of breaking reliance on predefined mutators, but \tool{} focuses on leveraging structural relations to test optimization logic. Creal only generates C programs, while \tool{} is language-agnostic. In addition, using Creal requires a large function dataset (50k+ extracted from Github in its evaluation). \tool{} only needs a small optimization corpus (100 programs) to derive its mutators.  

\noindent {\bf Fuzzing compiler optimizations.}
There exists a group of compiler fuzzers that are dedicated to, or are especially suitable for fuzzing optimizations. WhiteFox~\cite{yang2024whitefox} also leverages LLMs and use optimization-triggering tests for few-shot prompting. Thus, they steer testing towards compiler optimization logic. Optimuzz~\cite{kwon2025pldiOptimuzz} uses directed fuzzing to test newly-updated optimization code, while using translation validation for detecting miscompilations~\cite{lopes2021alive2}. It relies on some basic mutators that preserve the control-flow, limiting its exploration space. MopFuzzer~\cite{xie2025mopfuzzer} crafted several mutators that invoke optimizations, e.g, replacing an expression with a trivial function call that returns itself, in hopes of invoking function inlining. MopFuzzer is built for 13 specific optimizations in JVM's JIT compiler, while \tool{} is more general and can be applied to any optimizations in any compiler. CTOS~\cite{Jiang2022CTOS}, to the best of our knowledge, is the only optimization fuzzer other than \tool{} that also explicitly emphasizes and addresses the phase-ordering problem. CTOS uses vector embeddings of test programs and optimization sequences to cluster and prune similar instances, increasing test diversity. \tool{} mines and reconstructs grammar-based composition styles to directly test each specific compiler optimization with a targeted pipeline harness. The EMI family~\cite{le2014compiler, sun2016EMILiveCodeMutation, jiang2020cudasmith} of compiler fuzzers are also applicable to fuzzing optimizations by performing semantics-preserving transformation to obtain free miscompilation-oracles. 
\vspace{-.5em}
\section{Conclusion}
In summary, this paper is the first to {repurpose grammar-based fuzzing with composition styles to carry out targeted fuzzing of compiler optimizations}. We propose \tool{}, a general-purpose fuzzer that can be adapted to fuzz different compilers and programming languages at a low cost. Our evaluation shows that the success of composition-based mutators generalizes to a wide range of optimizations in LLVM and MLIR and that targeted fuzzing complements the existing compiler fuzzing practice.

\bibliographystyle{ACM-Reference-Format}
% \bibliography{reference}

\end{document}